%% file: main.tex
\documentclass[%
 superscriptaddress,
 amsmath,amssymb,
 prx,
 twocolumn,
]{revtex4-2}

\usepackage{hyperref}

\usepackage{cleveref}
\crefname{figure}{Fig.}{Figs.}
\Crefname{figure}{Fig.}{Figs.}
\crefname{equation}{Eq.}{Eqs.}
\Crefname{equation}{Eq.}{Eqs.}
\crefname{appendix}{Appendix}{Appendices}
\Crefname{appendix}{Appendix}{Appendices}
\crefname{section}{Sec.}{Secs.}
\Crefname{section}{Sec.}{Secs.}

\usepackage{graphicx}
\usepackage{amsmath}
\usepackage{amssymb}
\usepackage{braket}
\usepackage{xcolor}
\definecolor{lightred}{rgb}{0.9,0.4,0.4}
\definecolor{lightgreen}{rgb}{0.4,0.9,0.4}
\definecolor{lightblue}{rgb}{0.4,0.6,1.0}
\definecolor{mystarred}{HTML}{B62626}
\definecolor{mystarblue}{rgb}{0.1,0.3,0.7}
\newcommand{\red}[1]{{\color{lightred}#1}}
\newcommand{\green}[1]{{\color{lightgreen}#1}}

\newcommand{\nn}{\nonumber}
\usepackage{natbib}
\begin{document}

\title{A fidelity metric for quantum annealing benchmarked by extreme scaling quantum Monte-Carlo simulations}

\author{Gabriel Gouraud}
\email{gab.gouraud@gmail.com}
\affiliation{Univ. Grenoble Alpes, CEA, Grenoble INP, IRIG, Pheliqs, F-38000 Grenoble, France}
\author{Miha Srdinšek}
\affiliation{Univ. Grenoble Alpes, CEA, Grenoble INP, IRIG, Pheliqs, F-38000 Grenoble, France}
\author{Xavier Waintal}
\affiliation{Univ. Grenoble Alpes, CEA, Grenoble INP, IRIG, Pheliqs, F-38000 Grenoble, France}
\date{\today}

\begin{abstract}
Quantum annealers are supposed to follow adiabatically the ground state
of a system as its Hamiltonian slowly interpolates between a trivial phase
and a non-trivial one; the non-trivial ground state being the solution to an optimization problem. Overwhelmingly, their performances are measured in terms of how well or fast the optimization problem is solved.
While pragmatic, this approach is inherently brittle as it  strongly depends on the problem considered and the classical algorithm used as the reference benchmark. Here, we propose a quantity that not only measures  the end result but also the quality of the actual quantum annealing process itself. Our metric is the quantum annealing counterpart of the fidelity-per gate of
gate-based quantum computers. It takes the form of an accuracy $\epsilon$ for the equation of state of the annealer. We calculate benchmark values of $\epsilon$ using two variants of the simulated quantum annealing technique for Rydberg atoms systems. 
Our first approach uses variational quantum Monte-Carlo
with an ansatz inspired by thermal annealing. It suggests that within $\epsilon \sim 10^{-2}-10^{-3}$, a quantum annealer is indistinguishable from its thermal classical counterpart. 
Critically, we could reach this precision up to $100,000,000$ atoms on a single CPU. Our second approach (based on Green function quantum Monte-Carlo) reaches accuracies around $\epsilon \sim 10^{-4}$ and we have run it up to $100,000$ atoms. These results outperform current Rydberg atom quantum annealing experimental platforms in both precision and size by orders of magnitude and put severe constraints for future hardware.
\end{abstract}

\maketitle

\input{sections/introduction}
\input{sections/qa_review.tex}

\input{sections/thermal_ansatz.tex}
\input{sections/results.tex}

\section{Conclusion}
Difficulties seldom go away because one stops looking at them.
In this instance, we believe that the role of noise or decoherence must be understood if quantum annealing is to move forward further.
In this article, we raise this question: how the quality of a quantum annealer qubits is linked to its ability to solve the associated QUBO problem. While we do not answer the question directly, we propose a metric to measure how well a given quantum hardware (or simulation algorithm) performs its proposed task, quantum annealing. We argue that this is more informative than merely looking at whether the QUBO problem is solved or not. In particular, the experimental study of the accuracy of the equation of state should allow us to differentiate intrinsic limitations (diverging annealing time) from extrinsic ones (the current level of noise in a given hardware).

A secondary finding of this article is that a deceptively simple variational ansatz, inspired by thermal annealing, can lead to fast and relatively accurate simulations of QA. Our benchmark achieves an accuracy of $\epsilon \sim 10^{-4}$ for a $100,000$ Rydberg atom system. This research direction could be pushed further. For instance, we could relax entirely the constraints on $J_{ij}^\text{eff}$ or even consider three-body interaction terms or even higher order terms. This is a form of classical to quantum mapping that is perhaps unusual. Conceptually, it amounts to considering quantum annealing as the thermal annealing of a more expressive problem. This sort of analysis could help us to identify regimes where quantum annealing could bring a significant advantage with respect to genuine thermal annealing (beyond its simulation version with restricted Monte-Carlo moves) or, perhaps, help to build new heuristics to solve discrete optimization problems directly.

 \section*{Acknowledgements}

Warm thanks to A. Browaeys for interesting discussions and to L. Henriet 
and L. Leclerc for sharing the QUBO instances of Ref.~\cite{leclerc2022financial}.
X.W. acknowledges
funding from Plan France 2030 ANR-22-PETQ-0007
“EPIQ”, the PEPR “EQUBITFLY”, the ANR “DADI”,
the ANR TKONDO and the CEA-FZJ French-German
project AIDAS.

\appendix

\input{sections/appendix_beta.tex}
\input{sections/appendix_edge.tex}
\input{sections/appendix_perturbation.tex}

\input{sections/appendix_gap.tex}

\input{sections/appendix_vscore.tex}

\input{sections/appendix_range.tex}

\input{sections/appendix_first_order.tex}
\bibliography{bibliography/references.bib}

\end{document}

%% file: sections/introduction.tex
\section{Introduction}
Among the many possible applications that have been proposed
for quantum computing, discrete optimization problems would appear to be a
natural case.
Indeed, even though the internal state of an $N$-qubits quantum computers
lives in an exponentially large Hilbert space of $2^N$ dimensions, at the end of the calculation
one only gets a very small amount of information, merely $N$ classical bits.
It is therefore imperative that these $N$ classical bits be highly significant. Discrete optimization problems fit that particular bill: finding the most likely time for, say, the next financial crash, may lead to important economies and is therefore definitely worth the effort.
These problems are general, with applications spanning a vast range of contexts, from optimizing railway networks and designing school schedules to minimizing portfolio risk and navigating traffic.

For the purpose of this article, we suppose that the optimization problem has been cast in the form of a Quadratic unconstrained binary optimization (QUBO) problem \cite{kochenberger2014}: given $N$ bits $n_1\dots n_N$ and a symmetric matrix $J_{ij}$, one seeks the sequence of bits that minimize the functional
\begin{equation}\label{eq:qubo}
F(\mathbf{n}) = \sum_{ij} n_i J_{ij} n_j = 2\sum_{i<j} n_i J_{ij} n_j + \sum_{i} J_{ii} n_i,
\end{equation}
where the second form stems from $n_i^2 = n_i$. QUBO encompasses a wide class of optimization, including the NP-hard problems such as the traveling salesman or satisfiability SAT problems.  $J_{ij}<0$ favors sequences which satisfy ($n_i$ AND $n_j$) while using $J_{ij}>0$ with additional diagonal terms $J_{ii}=J_{jj}=-J_{ij}$ favors sequences which satisfy ($n_i$ XOR $n_j$). By combining different terms, possibly with the use of ancilla bits, one can express all kinds of constraints that need to be satisfied by the sequence of bits.

The idea of quantum annealing~\cite{kadowaki1998,morita2008,ohzeki2011} is to initialize the system in a regime where its Hamiltonian is trivial and its known ground state can be prepared efficiently. Then, the Hamiltonian is slowly changed into another one whose ground state we are interested in. If the evolution is adiabatic, the system should remain in its ground state and thereby prepares the desired target state. In the present context, the trivial Hamiltonian corresponds to a Zeeman magnetic field along the $x$-direction
\begin{equation}
\hat H_\text{x} = \sum_i \sigma_i^\text{x}.
\end{equation}
Here $\sigma_i^\text{x}$, $\sigma_i^\text{y}$ and $\sigma_i^\text{z}$ are the Pauli matrices applied on qubit $i$.
The Hamiltonian of interest is
\begin{equation}
\hat H_\text{z} = \sum_{ij} J_{ij} \hat n_i\hat n_j = \sum_{\textbf{n}}\  F(\mathbf{n}) |\mathbf{n}\rangle\langle \mathbf{n}|
\end{equation}
with the occupation operator defined as $\hat n_i = (1+\sigma_i^\text{z})/2$. The total annealing Hamiltonian is a transverse field Ising model and reads
\begin{equation}\label{eq:ising}
\hat H = \hat H_\text{z} - h_\text{x} \hat H_\text{x}.
\end{equation}
The quantum annealing protocol consists of (i) setting a high value of $h_\text{x}$; (ii) preparing the system in state $|++\dots +\rangle$ with $|+\rangle = [|0\rangle+|1\rangle]/\sqrt{2}$; (iii) slowly reduce $h_\text{x}$ towards $h_\text{x}=0$; (iv) measure the qubits to obtain a configuration
$\mathbf n=(n_1\dots n_N)$ that should be the solution of the QUBO problem.

Quantum annealing (QA) has been studied for the whole scope of applications that
can be formalized as QUBO problems. The seemingly limitless list of examples includes  many logistics problems \cite{phillipson2025}
(i.e., network design, scheduling, and routing), air traffic management \cite{Stollenwerk2021}, financial portfolio optimization \cite{Venturelli2022}, generative AI \cite{Benedetti2017} and the prediction of financial crashes \cite{orus2019}; see Refs.~\cite{yarkoni2022,smith-miles2025} for recent reviews. With such a putative impact, it is no wonder that quantum annealing has raised such a considerable interest. Hardware-wise, the two principal platforms are based on superconducting circuits \cite{King2018,King2021,King2022} and Rydberg atoms \cite{scholl2021,ebadi2021,ebadi2022,scotti2024} on which we focus in the present article.
The question at the center of all these works is, of course, what advantage will QA bring to solve these problems.

In the leading quantum computing approach, gate-based quantum computing, the central metric that determines what the hardware can and cannot do is the so-called gate-fidelity. Getting another ``9'' in the fidelity (e.g., from 99\% to 99.9\%) takes years of drastic efforts with little guarantees of success. The entire program of ``fault tolerant quantum computing'' followed by the leading quantum companies has the unique goal of increasing this number through the implementation of quantum error correction. It may therefore come as a surprise that the question of the fidelity seems to have evaded completely the conversation in QA. Most works take a very pragmatic perspective: let's just try and see if QA actually solves the problem more efficiently than its classical counterparts. However, this capability has not yet been demonstrated, and important theoretical objections remain, including exponential slowdowns on hard instances due to Landau-Zener transitions. We therefore argue that directly measuring the extent to which quantum annealing performs its intended task is a crucial step forward. In this work, we propose to use the accuracy of the QA equation of state $E(h_\text{x})\equiv \langle\hat H\rangle$
\emph{during} the annealing. If decoherence, finite-temperature effects, or finite annealing times prevent an accurate reconstruction of the equation of state, these errors are unlikely to compensate and restore the accuracy of the final result. On the other hand, the accuracy of the reconstructed equation of state provides a direct diagnostic of the dominant mechanisms responsible for deviations from ideal adiabatic evolution.

As a benchmark for our proposed metric, we perform two kinds of \emph{classical} calculations. The first is a variational Monte-Carlo (VMC)
calculation with an ansatz that bears strong similarities with thermal annealing. This somewhat twisted approach -- using thermal annealing of an altered classical system to simulate a quantum annealer -- will provide key insights as to the differences and similarities between the fluctuations
provided by quantum mechanics and those provided by mere thermal excitations. We show calculations of up to $100,000,000$ atoms on a single CPU with an estimated accuracy of $\epsilon\sim 10^{-2}-10^{-3}$ for Rydberg atoms.
The second is a Green-Function Monte Carlo (GFMC) calculation that provides a more accurate result at roughly the same computing price. We show calculations of up to $100,000$ atoms with an estimated accuracy of $\epsilon\sim 10^{-4}$ for Rydberg atoms. These results suggest that it will be very challenging for these Rydberg QA to beat classical approaches.

The rest of the article is organized as follows. Section~\ref{sec:qa_review} gives a critical overview of what is QA. We discuss both its theoretical limitations, its main competitors, and the previous experiments with superconducting qubits and Rydberg atoms.
Section~\ref{sec:vmc} introduces our approach to describing QA with a form of thermal annealing.
In the last two sections, we test our approach.
In Section~\ref{sec:sim} we consider Rydberg atoms on regular lattices.
In Section~\ref{sec:qubo}, we consider a real-life QUBO instance associated with the ``fallen angel problem''.

%% file: sections/qa_review.tex
\section{Can quantum annealing give a speed-up?}
\label{sec:qa_review}

In this section, we briefly review a few claims that have been made about QA
and try to clarify what can be reasonably expected from the technique, from
a general perspective. We will not discuss in any detail the various technical limitations of the hardware. Those include the presence of noise, relaxation or decoherence in the hardware, which $J_{ij}$ matrix is supported  (which implies embedding strategies with their associated overheads), the difficulties associated with scaling to large system sizes, etc. All these difficulties are platform dependent. While we will not study them, we will propose a metric to study their impact.

\subsection{An exponential annealing time}
For an ideal QA, the main limitation is the fact that one uses a finite
 total annealing time $t_\text{tot}$ to bring $h_\text{x}$ from its initial value down to zero.  Hence, the relevant parameter is the speed
$\dot h_\text{x} \equiv dh_\text{x}/dt \sim h_\text{x}(t=0)/t_\text{tot}$ at which the field is decreased. This velocity $\dot h_\text{x}$ controls the corrections to adiabaticity, i.e. the probability that one does not end up in the ground state after the annealing process. This problem has been studied extensively and is known as the Landau-Zener problem \cite{zener1932}. The
simplest situation is a $2\times 2$ problem with a Hamiltonian given by
$\hat H = \Delta \sigma^\text{z} + \dot h_\text{x} t \sigma^\text{x}$ and an initial state
$|\Psi(t=-\infty)\rangle = |+\rangle$. The probability not to be in the
ground state at $t=\infty$ (which is $|-\rangle = [|0\rangle-|1\rangle]/\sqrt{2}$) is given by
\begin{equation}
 P_\text{exc} = \exp\left(- \frac{\pi \Delta^2}{2 \hbar |\dot h_\text{x}|}\right).
 \label{eq:LZ}
 \end{equation}
It follows that, in order for the annealing to succeed ($P_\text{exc}\ll 1$) the annealing time must scale as the square of the inverse of the gap $\Delta$ between the ground state and the first excited state,
$t_\text{tot} \gg \hbar h_\text{x}/\Delta^2$. The success of quantum annealing is therefore entirely conditioned by the minimum of the gap $\Delta(h_\text{x})$ of $\hat H(h_\text{x})$ as $h_\text{x}$ is reduced.

Unfortunately, there seems to be a conspiracy of Nature: problems with large gaps (where QA can succeed) tend to be easy to solve classically.
For instance, it can be proven rigorously that gapped local Hamiltonians in one dimension satisfy the area law for entanglement and can therefore
be simulated in polynomial time with tensor networks \cite{white1992,wolf2008}.
Conversely there is a strong accumulated evidence that hard QUBO problems have exponentially vanishing gaps which implies exponentially long annealing time. This problem has been studied in a number of situations such as spin glasses, random transverse fields and more \cite{miller1993,vandam2001,reichardt2004,jorg2008,amin2009, altshuler2010, young2010,jorg2010, farhi2011,miyazaki2013,bapst2013}. Many
(but not all \cite{bapst2013}) of these systems display a first order transition with an exponentially small gap;
however the connection between the two is not fully established \cite{laumann2012}.

The situation therefore is in sharp contrast with, say, Shor's algorithm where quantum computers are expected theoretically to bring an exponential speed-up: in QA we do not have generic reasons to expect any. It is therefore unlikely that QA will prove transformative for solving QUBO. It remains however entirely possible that some advantage can be gained in practice for some problems in some regimes.

\subsection{The competition to quantum annealing}

Depending on one's point of view and goal, there are different competitors to QA that one may compare to. For a given concrete application and
a pragmatic point of view, one wants to compare to the best existing classical solver which is often a heuristic that has little or nothing to do with quantum physics. However, if one is interested in understanding what the "quantum" in QA brings to the computing capability then considering its classical counterpart (thermal annealing) or its simulated counterpart
(simulated quantum annealing) may be relevant too. In this section, we briefly discuss these different options successively.

\subsubsection{Thermal annealing}
The most natural counterpart to QA is to build a classical machine
tailored to solve QUBO problems using thermal fluctuations instead of
quantum ones. This route is becoming increasingly popular with, e.g., demonstrations of up to $2000$ artificial spins in photonic systems \cite{inagaki2016} with all-to-all connectivity. A large number of different technologies may be used, from spintronics to memristors or even conventional semiconductors, see \cite{mohseni2022} for a review.

Indeed, a quantum annealer merges two different ideas. The first is to build a dedicated machine tailored to solve discrete optimization problems. The second is to use quantum fluctuations. Given the important resources needed to scale and operate a quantum annealer, it is indeed relevant to wonder which of the former or the latter contributes more to its performances.
Note that nature is not necessarily better at solving QUBO problems than us:
in glasses, for instance, the thermalization time of the system diverges as one
approaches the glass transition.

\subsubsection{Simulated thermal annealing}
The baseline to which many QA articles compare themselves is a simulated version of thermal annealing, often simply referred as simulated annealing (SA). In SA, one uses a Markov Chain to sample a thermal distribution $P(\mathbf{n})= (1/Z) \exp[-\beta_\text{eff} F(\mathbf{n})]$
and then gradually increase the inverse temperature $\beta_\text{eff}$ until one has reached the ground state of the QUBO problem. It is a very generic and old technique \cite{kirkpatrick1983,vcerny1985,hajek1988} that contains many variants. The art of SA, and Monte-Carlo techniques in general, is to build sampling strategies that change many spins at once (so-called global updates) using either physics-based heuristics or more complex constructions (e.g. quantum cluster updates \cite{sandvik2003}). Most comparisons with QA use simple local updates which is often suboptimal. In SA, the hardness of the QUBO problem manifests itself in the difficulty for the Markov chain to be ergodic and therefore equilibrate without getting trapped in local minima.

\subsubsection{Simulated quantum annealing}

Simulated quantum annealing (SQA) is a generic terminology that encompasses various quantum Monte-Carlo techniques that aim at simulating the ground state of the QA problem itself. Importantly, the QA Hamiltonian is sign-problem \emph{free} so that these techniques are in principle exact. However, SQA potentially suffers from the same problem as QA, an exponentially long annealing time in the presence of frustration. Most comparisons between QA and SQA have been performed with Path Integral Monte-Carlo (PIMC) \cite{ceperley1995}, a popular variant. As in SA, the performance of quantum Monte-Carlo techniques strongly depends on the sampling strategy.

SQA has been proposed as an alternative route to SA to solve QUBO problems
as early as \cite{kadowaki1998,morita2008,ohzeki2011}. It has been extensively used as a proxy of QA to look for an advantage of QA over SA.
In the literature, SQA has been used for several distinct purposes that should not be confused:
(i) to identify specific features of QA with respect to SA in order to demonstrate that quantum mechanics is really at play in an experiment;
(ii) to demonstrate the advantage of SQA over SA and therefore the putative advantage of QA over SA; (iii) more rarely, to contrast SQA with QA in order to provide arguments in favor of doing QA rather than merely simulating it on a classical computer.

Some scaling advantage of SQA over SA, in the sense of the residual energy
dropping faster as a function of a proxy of the annealing time, was observed in spin glass \cite{santoro2002}. However, the advantage was later found to be present only with discretized imaginary time, not in the continuum
\cite{heim2015}, hence likely to be present in SQA but not in QA. A claim of superiority of SQA over SA can also be found for the traveling salesman problem \cite{martonak2004}. Other authors have carefully crafted problems in order for SQA to be provably better than SA, designing energy landscape with local minima that have a large basin of attraction and exponentially small
global minimum \cite{farhi2011}. Ref.~\cite{vandam2001} constructed two kinds of instances: some where SQA is advantageous and some where SQA and QA provably take an exponentially long time to find the solution. It should be noted, however, that exponentially small gap may depend on the annealing path since adding extra terms in the Hamiltonian may allow one to bypass certain first order transitions \cite{seki2012}.

\subsubsection{Classical heuristics}

Many QA applications originate from groups that had a history in
solving discrete optimization problems, long before QA became fashionable.
While the worst case solution of many of these problems is provably NP hard, hence almost certainly exponentially hard, practical instances have a lot of extra structure that specialized algorithms may take advantage of for considerable speed-ups. These heuristic algorithms are often the state of the art used in production for real applications. Overwhelmingly, the conclusions of these studies are that current implementations of QA are not yet competitive with existing heuristics. Reviews can be found, e.g., in \cite{yarkoni2022}; a specific review is devoted to the emblematic example of the traveling salesman problem and is rather pessimistic including for future
prospects \cite{smith-miles2025}. Perhaps more optimistic is \cite{quinton2025} for \emph{some} specific applications, but this study should be taken with care because it uses a "hybrid" approach where
classical heuristics with CPUs/GPUs are involved also on the quantum side, hence determining the contribution of QA in the performance is non trivial.
Ref.~\cite{carugno2022} studied a class of difficult instances of the
job shop scheduling problem. There, it was found that SA was superior to QA for large instances.

For Rydberg atoms, Ref.~\cite{serret2020} estimated that a few thousand atoms would be needed (together with improvement on the noise level) for obtaining an advantage for the Unit-Disk Maximum Independent Set problem. More recent estimates \cite{cazals2025} seem to give the same order of magnitude.

\subsubsection{Other classical baselines}

The examples above do not exhaust the classical approaches against which quantum annealing can be compared. As the importance of discrete optimization problems spreads, new classes of classical algorithms are likely to emerge.
Examples include a combination of belief propagation with a tensor network
\cite{luchnikov2024} (Ref.~\cite{luchnikov2024} also contains a comprehensive comparison with many heuristics).
Neural networks either in the form of large language models \cite{qian2026} or as variational ansatz for simulating quantum annealing \cite{mauron2025}
are also obvious candidates.

Ref.~\cite{shin2014} used a semi-classical mapping of the qubit seen as a spin-$1/2$. The simple, yet elegant, idea is to map the
quantum spin to a classical 2D spin model, hence to a form of SA but in a larger -- continuous, not discrete -- space. The authors found that the results of these perfectly classical simulations were very close to what was observed in SQA, thereby raising doubts on the validity of some experimental signatures used to indicate quantumness.

\subsection{Selected quantum annealing experiments}

\subsubsection{Superconducting circuits}

The first implementation of QA with superconducting circuits is
now already fifteen years old \cite{johnson2011}. It developed into roughly three stages.

Initially, the focus was to demonstrate that there actually was some quantumness involved in the experiments. Ref.~\cite{lanting2014}
prepared a superposition of the two ferromagnetic states of a small system with up to $8$ qubits (a GHZ state) thereby showing some evidence of entanglement. Scaling up to $\sim 100$ qubits, Ref.~\cite{boixo2014} showed evidence that the experimental results were better described by SQA than SA.
However, this evidence of quantumness was deemed insufficient later by Ref.~\cite{shin2014} who exhibited an SA-like method consistent with the SQA one.
Scaling up to $\sim 1000$ qubits, Ref.~\cite{King2018} switched from QA to
a quantum simulation mode and built an effective model that described
a Kosterlitz-Thouless phase transition. Again, the experimental data were resembling those of a corresponding SQA simulation.
All these experiments clearly showed that while/if quantum mechanics was involved, thermal effects (included in PIMC SQA simulations) clearly played a significant role, with a temperature equal to a fraction (say $\sim 1/4$) of the exchange energy $J_{ij}$. Ref.~\cite{King2023} performed time resolved experiments and showed that the coherent theory was consistent with the experimental results when the annealing was done rapidly (in less than $100$ns). Note that this is the opposite limit of annealing, almost a quench,
so it is not expected to be useful for solving QUBO problems. In the most recent experiment, Ref.~\cite{king2025} used such a quench to claim a "quantum advantage", i.e. claiming that it would be intractable for a classical method to simulate the experiment. However, the claim was refuted almost immediately by Ref.~\cite{mauron2025} who used the time dependent variational principle implemented on a neural network to simulate the experiment.

In the second stage, the quest for a QA advantage involved building tailored problems suitable for QA but not for SA. Ref.~\cite{denchev2016} found that
QA was orders of magnitude faster than SA. However, this result came with two important caveats. First, it relied on the highly specific Chimera graph, the native qubit connectivity of the device, which is not representative of typical QUBO problems and
 for which there are classical PTAS (provably good approximate algorithms that run in polynomial time) \cite{saket2013}.
Second, the energy landscape was crafted to have tall and narrow barriers
where tunneling is efficient but thermal annealing is not. Again, it is not clear if such planted features are present in some real QUBO instances.
Also, the speedup was only present relative to SA, not with respect to the best classical heuristics. Ref.~\cite{albash2018} also engineered tall and narrow barriers and found better performances than SA but not as good as SQA.
The latest of this series is Ref.~\cite{King2021} which used a topological obstruction and found better performances than SQA.
However, we note that the PIMC simulation used only local updates; more global updates might get past the obstruction. This experiment was still limited by the rather high temperature mentioned above which put large constraints to resolving small energy scales.

The third stage was about applying the QA approach to industrially relevant problems. It was recognized that the relatively small connectivity of the qubits (a few percent of all possible couplings are present) needed to be addressed as it created overheads both along space (several qubits need to be used per site of the QUBO problem) and time (to calculate the correct embedding classically \cite{carugno2022}) and could even change the overall scaling of the approach \cite{konz2021}. Ref.~\cite{vert2021} considered instances of the bipartite matching problem known to be hard for SA but they turned out to be hard for QA too. Ref.~\cite{mohseni2022} considered two  problems with dense connectivity (Sherington-Kirkpatrick and maxCut) and observed a collapse of the QA success probability for large instances. Overall, most articles of this third stage (with the notable exception of Ref.~\cite{smith-miles2025}) have an optimistic tone, but point to existing bottlenecks.

\subsubsection{Rydberg atoms}

Rydberg atoms QA platforms are more recent than superconducting circuits and the corpus of studied applications is much smaller. Each qubit $i$ is an atom that can be placed  at position $\mathbf{r}_i$ with optical tweezers
resulting in an interaction matrix
\begin{equation}\label{eq:rydberg_hamiltonian}
J_{ij} = \frac{(1-\delta_{ij})}{2| \mathbf{r}_i - \mathbf{r}_j|^6} - \delta_{ij}h_\text{z},
\end{equation}
where $\delta_{ij}$ is the Kronecker symbol. The off-diagonal part results from a van der Waals interaction while the diagonal term, often spatially independent, operates
as a global Zeeman field.
The typical time available for the QA is of the order of $t_\text{tot}\leq 10\mu s$, currently limited by the drift of the atoms when the optical tweezers are removed. It could potentially be extended by around one order of magnitude where it would become limited by the intrinsic decay time of the Rydberg state. On the other hand, $J_{ij}$ itself is of the order of a few MHz, it follows that the systems can remain approximately adiabatic for $\Delta\sim J_{ij}$ but situations where $\Delta\ll J_{ij}$ would require a different approach.

Experiments in a quantum simulation mode showed elements of the antiferromagnetic and ferromagnetic phase diagram for several $\sim 100$ atoms \cite{scholl2021,ebadi2021} and the QA mode was demonstrated for
the maximum independent set problem (which does not require embedding) in
Ref.~\cite{ebadi2022}. Examples of industrially relevant problems include
Ref.~\cite{leclerc2022financial} to which we shall come back below; see \cref{sec:qubo}.

\subsection{Discussion of the methodology}

Overwhelmingly, the methodology that has been used in the literature is
to compare the performance of QA to one or more classical methods, being SA, SQA or a heuristic, for solving certain instances of QUBO. This is perfectly adequate for evaluation of current QA hardware, but these results are intrinsically brittle as they strongly depend on the problem, on its size and on the
classical method (including tiny crucial details such as the update strategy in SQA). The lack of understanding of the source of a failure -- intrinsic
(vanishing gap) versus extrinsic (i.e. noise, finite temperature) -- makes it difficult to evaluate possible future prospects.

In this work, we contribute to the question by slightly shifting the discussion: instead of trying to solve the QUBO problem itself, we propose to evaluate how the quantum annealer actually performs quantum annealing.
We argue that a small change in the annealing protocol would achieve this goal: for a given targeted final value of $h_\text{x}$ (that is not necessarily $h_\text{x}=0$ in contrast to the QUBO mode), one would need to make two sorts of experimental runs: normal runs where the qubits are measured along the $Z$ axis (this provides the value of the Ising energy $\bra{\Psi} \hat h_\text{z} \ket{\Psi}$) and a second sort of runs where a Hadamard gate is applied at the end of the annealing so that the measurement is performed along the $X$ axis (this provides the value of the transverse field energy $\bra{\Psi} \hat H_\text{x} \ket{\Psi}$). Together, these two contributions give the equation of state,
\begin{equation}
E(h_\text{x}) \equiv \bra{\Psi} \hat H \ket{\Psi}
\end{equation}
of the system.
More precisely, we are interested in the accuracy with which one can obtain this equation of state,
\begin{equation}
\epsilon = \delta E/E.
\end{equation}
We argue that this accuracy is a good proxy for a quantum annealing fidelity.
First, the equation of state contains most of the physics as its derivatives describe the order parameters and correlation functions. Second, it will provide information as to the different sources of errors: for instance the value of $\epsilon$ away from a gap-closing will tell about the best accuracy one can hope for, even for easy QUBO problems, while the value around a gap closing will quantify the more intrinsic problem of Landau-Zener transitions that will appear for hard instances.
Third, classical simulations of the equation of state will provide a good benchmark of the level of accuracy needed to challenge classical methods, in the sense that if the equation of state can be
simulated with precision, it implies that the QUBO problem can be solved with similar precision.
The rest of this article is devoted to providing the said classical benchmark for the case of Rydberg atoms.

%% file: sections/thermal_ansatz.tex
\section{How quantum is quantum annealing: a variational approach.}
\label{sec:vmc}

In the early development of quantum annealing, a central question was whether and how it could be distinguished from thermal annealing.
 Clearly, on physical grounds, quantum and thermal fluctuations are different. For instance, quantum tunneling can provide a path through an extremely high potential barrier provided it is also extremely narrow, while thermal fluctuations
cannot. Yet, quantifying how different the two types of process are in the specific case of QA is not straightforward. For instance, it was shown in Ref.~\cite{nishimori1996} that, in the case of a Hopfield model in a transverse field, the quantum fluctuations
are equivalent to thermal fluctuations. Ref.~\cite{Raymond2020} found the emergence of a local temperature in a QA.
Ref.~\cite{shin2014} studied a classical model which showed a behavior very close to those observed in a QA. A more mundane aspect of the question is that actual quantum annealers do have a finite temperature and it is not entirely clear to what extent this should be treated as a resource (of fluctuations) or a nuisance (to reach accuracies below the temperature).

To address this question -- and at the same time construct our first classical benchmark -- we will use a somewhat convoluted approach. We consider a variational
wavefunction
\begin{equation}
\label{eq:thermal_wf0}
 \ket{\psi_\text{V}} = \sum_{\boldsymbol{n}} \psi({\boldsymbol{n}}) \ket{\boldsymbol{n}}
\end{equation}
which we want to use to obtain an approximate equation of state for our QA Hamiltonian.
We shall use the standard Variational Monte-Carlo (VMC) approach \cite{becca2017} to perform the calculations. Nowadays, a typical model for $\psi({\boldsymbol{n}})$ would be a neural network
\cite{sprague2024} but tensor networks are also being used \cite{naumann_rizzi_introduction_2024} or combinations of the two \cite{srdinsek_2025}. Those models are very powerful and can get very close to the true ground state of the QA. An interesting property of VMC is that one measures directly the total energy of the state, as opposed to QA where one needs to measure the longitudinal and transverse components of the energy separately. It follows that a good variational ansatz (where the variance of the energy is low) requires very few samples to be accurate, see Ref.~\cite{louvet2026} for an in depth discussion.

Here, we consider the following ansatz
\begin{align}\label{eq:thermal_wf0_2}
\psi({\boldsymbol{n}}) =\frac{1}{\sqrt{Z}} e^{-E^\text{eff}(\boldsymbol{n})/2}
\end{align}
where we will choose an expression for the effective energy $E^\text{eff}(\boldsymbol{n})$ inspired from the classical energy $F(\boldsymbol{n})$. The factor $Z$ ensures the normalization of the state.
In plain words, we will use simulated thermal annealing on a modified system as an ansatz to simulate QA.
As advertised, this might seem somewhat convoluted; yet it will provide a direct measure of how close SA is to QA, \emph{quantified} by the obtained accuracy $\epsilon$. A secondary advantage of this ansatz is that it is very light computationally, enabling us to simulate very large system sizes on single CPU. Similar ansatz, in this context known as Rokhsar-Kivelson wavefunction, has been used in a reversed setting, where authors were proposing to use a quantum computer to emulate thermal states \cite{ardonne2004,henley2004,castelnovo2005,verstraete2006,tarabunga2024}. Here we do the opposite and map the quantum state onto a classical one (within an accuracy $\epsilon$).

More precisely, we can consider a hierarchy of models for $E^\text{eff}(\boldsymbol{n})$. The first is to use -- \emph{the} -- actual thermal annealing with an effective temperature
$\beta^\text{eff}$
\begin{equation}\label{eq:simple}
E^\text{eff}(\boldsymbol{n}) \equiv \beta^\text{eff} F(\boldsymbol{n}).
\end{equation}
This ansatz contains a single variational parameter whose optimum value will
depend on the transverse field $\beta^\text{eff}(h_\text{x})$. The corresponding simulation amounts to finding a temperature such that the thermal fluctuations are as close as possible to the quantum fluctuations.
We may, however, relax this assumption and look for
-- \emph{a} -- thermal annealing that describes the QA best. This second class of ansatz amounts to using
 \begin{equation}\label{eq:couplings}
E^\text{eff}(\boldsymbol{n}) \equiv \sum_{ij} J^\text{eff}_{ij}  n_i n_j,
\end{equation}
with a larger number of variational parameters $J^\text{eff}_{ij}(h_\text{x})$. We will consider both successively below. Note that we could further complexify the ansatz by adding e.g., three-body or even n-body interactions.
 The number of parameters in the ansatz can be reduced by considering the symmetries of the target quantum state (see \cref{app:beta}).
 The accuracy obtained with the ansatz Eq.~\eqref{eq:couplings} tells us about the advantage (or lack of) of building an actual thermal annealer to mimic QA. Technically, this thermal ansatz has two nice properties: first it is very light computationally and the derivatives
with respect to the effective temperatures are given in terms of simple observables without the need to use automatic differentiation. Second, it is trivially vectorial (both for ansatz evaluation and different walkers of the Markov chain) with a very low memory footprint. This would imply near perfect speedup (e.g., 100-1000) had we implemented the code on GPUs instead of the CPUs that were used in our numerics.

In our second benchmark, we will give up on the interpretability of the calculation and simply aim at having the most precise result at the lowest computational cost.
We will use Green Function Monte-Carlo (GFMC) for this purpose. GFMC is a post VMC method that evolves the variational ansatz in imaginary time in order to project it
(in the stochastic sense, i.e., to collect samples) onto the true ground state of the system. It provides an unbiased estimate for the energy
\begin{equation}
E(\tau) \equiv \bra{\psi_\text{V}} \hat H e^{-\hat H \tau} \ket{\psi_\text{V}}
\end{equation}
which converges to the equation of state at large time. GFMC does not suffer from any sign problem for the transverse field Ising models considered here.
Our implementations of both VMC and GFMC are described in \cite{srdinsek_2025}.

\begin{figure*}[htbp]
  \centering
  \includegraphics[width=\linewidth]{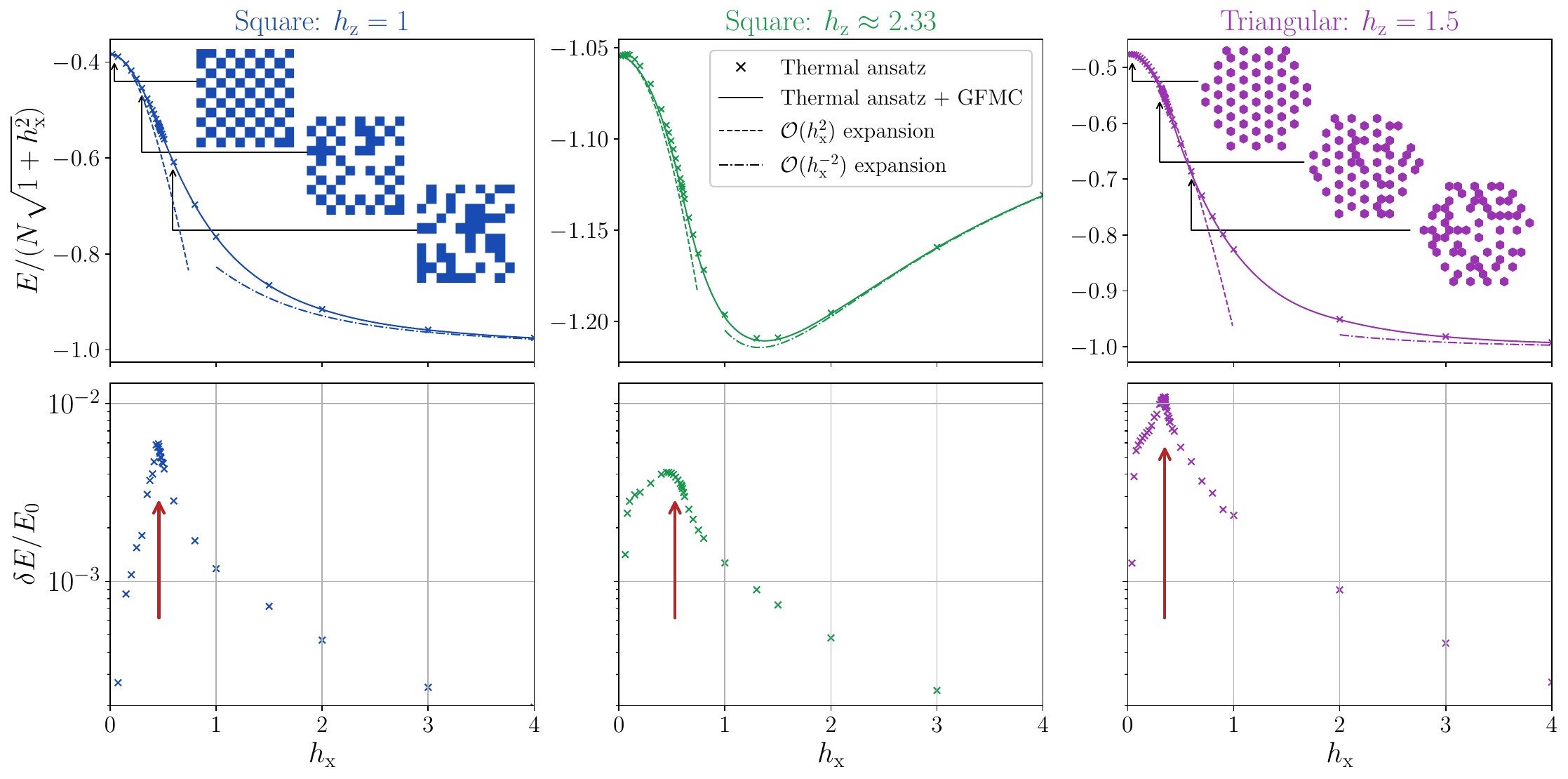}
  \caption{\textbf{Example of equations of state for the square and triangular lattices of Rydberg atoms.} Left: Square lattice for $h_\text{z} = -1$ and $N=100$ atoms. Middle: Square lattice for $h_\text{z}\approx2.33$ and $N=100$.
  Right: Triangular lattice for $h_\text{z}=1.5$ and $N=147$. Top panels: Energy as a function of transverse field computed with the staggered thermal ansatz Eq.~\eqref{eq:ansatz_stag} (cross),
staggered thermal ansatz plus GFMC (full line), second order perturbation theory in $h_\text{x}$ (dashed line) and in $1/h_\text{x}$ (dot-dashed line). The insets are snapshots of the Rydberg configurations at $h_\text{x}=0.02$, $h_\text{x}=0.3$
and $h_\text{x}=0.6$. Bottom panels: relative error of the thermal ansatz with respect to the GFMC result. The red arrows indicate the position of the gap computed using our GFMC gap method (see \cref{app:gap}).
  }
  \label{fig:eq_of_state}
\end{figure*}

%% file: sections/results.tex
\section{Benchmark of quantum simulations}
\label{sec:sim}

We start our benchmark with a system of Rydberg atoms \cref{eq:rydberg_hamiltonian} on two regular lattices: a square lattice and a triangular lattice.

\begin{center}
    \includegraphics[width=0.6\linewidth]{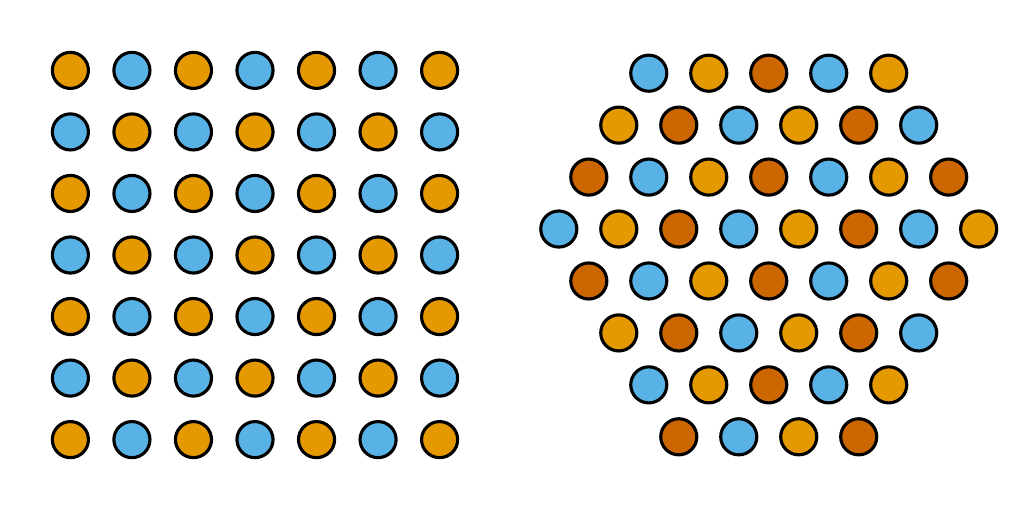}
\end{center}

The square lattice has $P=2$ sub-lattices
$\{A_0,A_1\}$ (orange and blue sites above) while the triangular lattice has $P=3$ sub-lattices $\{A_0,A_1,A_2\}$ (respectively orange, blue and red above). We define the average occupation of these sub-lattices and the average occupation of the lattice as
\begin{equation}
N_k(\mathbf{n}) \equiv \sum_{i \in A_k} n_i, \quad N(\mathbf{n}) \equiv \sum_k N_k(\mathbf{n}),
\end{equation}
and the corresponding quantum operator as $\hat N_k  \equiv \sum_{i \in A_k} \hat n_i$.

\subsection{Expected phase diagram and ansatz}
\begin{figure}[htbp]
  \centering
  \includegraphics[width=1.0\linewidth]{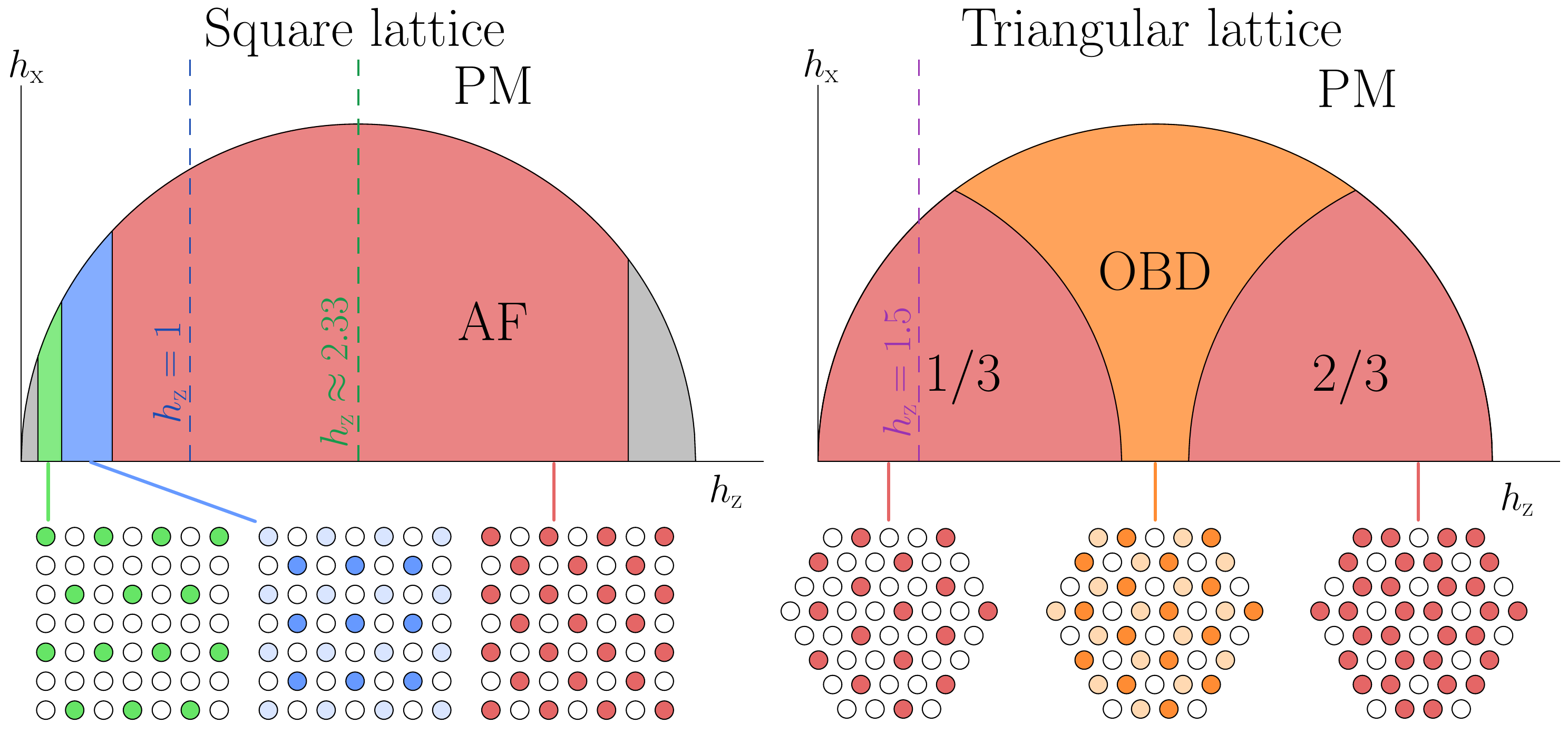}
  \caption{Sketch of the Rydberg atom phase diagrams for the square lattice (left) and triangular lattice (right); see \cite{kalinowski_2022}. The inset illustrates typical configurations (light blue and orange: partially occupied configurations).
  AF: anti-ferromagnet, PM: paramagnet, OBD: order by disorder,
  $1/3$ and $2/3$ anti-ferromagnet-like ordered phase.
 Vertical dashed lines: lines studied in Fig.\ref{fig:eq_of_state}.}
  \label{fig:phase_diag}
\end{figure}

The expected phase diagrams together with the three lines that we have studied are shown in Fig.\ref{fig:phase_diag}. The Rydberg Hamiltonian has a particle-hole symmetry around $h_\text{z} \approx -2.33$ (square lattice) and $h_\text{z} \approx 3.19$ (triangular lattice); see Appendix \ref{app:beta}. We focus in particular on the anti-ferromagnetic phases characterized by the order parameter (valid for both lattices),
\begin{equation}\label{eq:order_param}
M = \frac{P}{N} \left| \sum_{k} e^{2i\pi\tfrac{k}{P}}
\langle \hat N_k \rangle  \right|.
\end{equation}
Note that the Rydberg Hamiltonian has open boundary conditions. It follows that these systems display strong finite size effects with
a significant fraction of the atoms on the edges and corners,
see Appendix~\ref{app:deffect}.

The triangular lattice introduces geometric frustration and supports a richer phase diagram \cite{moessner1999,moessner2001,isakov2003,koziol2019,fey2019}. At $h_\text{z} \approx 3.19$
and $h_\text{x} =0$, the classical Ising model is strongly degenerate; an order by disorder  \cite{villain1980} (OBD) phase emerges as $h_\text{x}$ is switched on \cite{guo2023,bombieri2025}.
Below, we focus on the $1/3$-paramagnetic transition \cite{scholl2021}.

\subsection{Numerical benchmark.}
In the following, we have used two different ansätze to calculate the equation of state along the three
cuts shown in Fig.\ref{fig:phase_diag}: $h_\text{z}=-1$, $h_\text{z}\approx -2.33$ ($N=10\times 10$, square lattice)
and $h_\text{z} = -1.5$ ($N=147$, triangular lattice). The first and last cuts correspond to the experiments of Ref.~\cite{scholl2021}.

The two ansätze are Eq.~\eqref{eq:simple} (hereafter, the "simple thermal ansatz"), and a restriction of the ansatz Eq.~\eqref{eq:couplings} (hereafter the "staggered thermal ansatz")
with the following effective energy that contains simple symmetry breaking terms
 \begin{align}\label{eq:ansatz_stag}
E^\text{eff}(\boldsymbol{n}) = \beta^{\text{eff}}  F(\boldsymbol{n}) + \sum_k \beta_{k} N_k(\boldsymbol{n}).
\end{align}
These are very light ans\"atze with a maximum of just 4 variational parameters. Modern neural network ans\"atze can contain up to trillions of parameters for problems important enough to warrant the amount of computing time that is spent to train a large language model. For our small model, any gradient descent based algorithm very quickly converges to the optimum solution.

Fig.~\ref{fig:eq_of_state} shows the obtained equations of state
$E(h_\text{x})$ (upper panels)  and corresponding relative error
$\delta E/E_0$ (lower panels). At the scale of the upper panels, the results of the VMC simulations are almost indistinguishable from those of the GFMC which are themselves indistinguishable from
more accurate reference calculations that we have performed for specific points using the technique of \cite{srdinsek_2025}.
The small ($h_\text{x}\ll 1$, dashed) and large ($h_\text{x}\gg 1$, dotted dashed) fields parts of these equations of state can be understood from simple perturbation theory (see Appendix \ref{app:perturbation} for the explicit expressions) which are included to highlight the "difficult" regions of the phase diagram. Again, we argue that measuring these equations of states experimentally would provide a quantitative measure of how well
the quantum annealing process is performed. It essentially requires additional measurements in the x-basis compared with existing experimental data.

\begin{figure}[htbp]
  \centering
  \includegraphics[width=\linewidth]{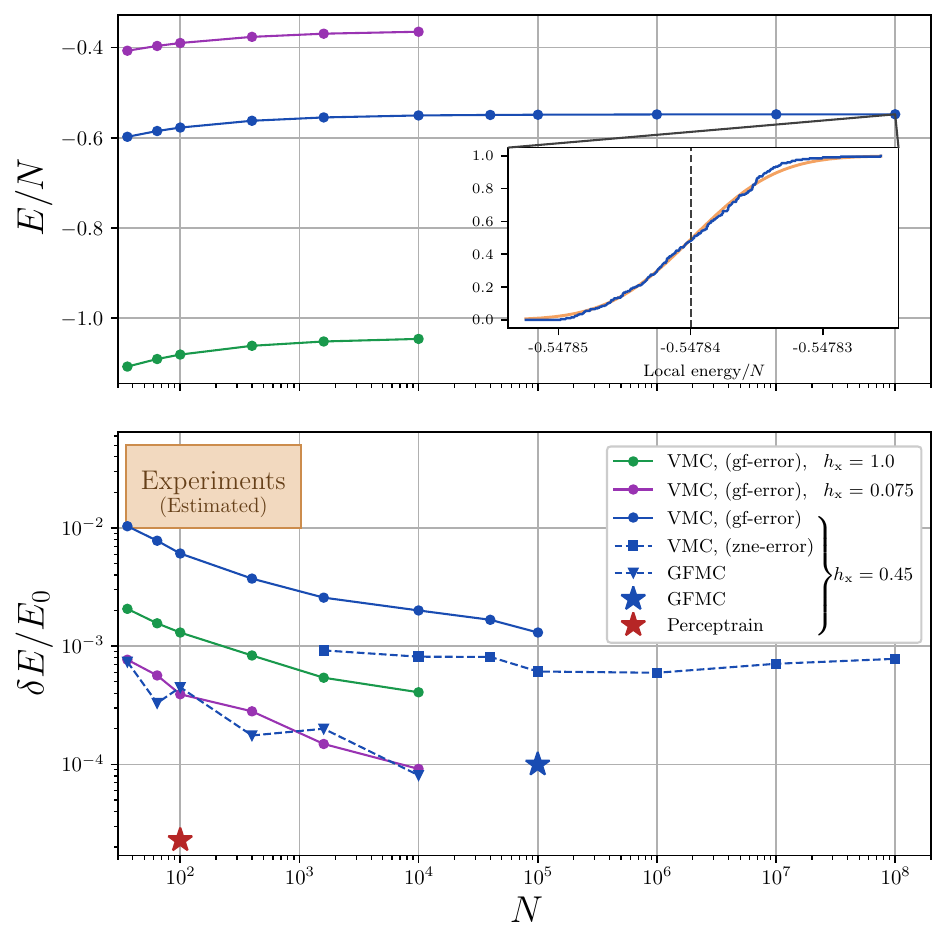}
  \caption{\textbf{Top panel: VMC energy $E/N$ versus system size $N$.} The circles correspond to the staggered thermal ansatz for
  the square lattice with $h_\text{z} = 1$, $h_\text{x}=1$ (green), $h_\text{x}=0.45$ (blue) and $h_\text{x}=0.075$ (purple).
 Inset: cumulative distribution of the local energy for the largest size $N=10^8$ (blue, 200 walkers) together with a fit with a gaussian distribution (orange).
\textbf{Bottom panel: error $\delta E/E_0$ versus system size $N$}: VMC error of the three curves of the top panel (circles, full lines as well as square and dashed line for the zne-error) and error of the GFMC calculation (only $h_x=0.45$ is shown, triangle with dashed line).  Blue \textcolor{mystarblue}{$\bigstar$} symbol: extrapolated error of the GFMC calculation, see text.
Red \textcolor{mystarred}{$\bigstar$} symbol: highly accurate calculation with a perceptrain ansatz, see text.
Orange rectangle: estimated capabilities of experiments.
   }
  \label{fig:scaling_error}
\end{figure}

\begin{figure}[htbp]
  \centering
  \includegraphics[width=\linewidth]{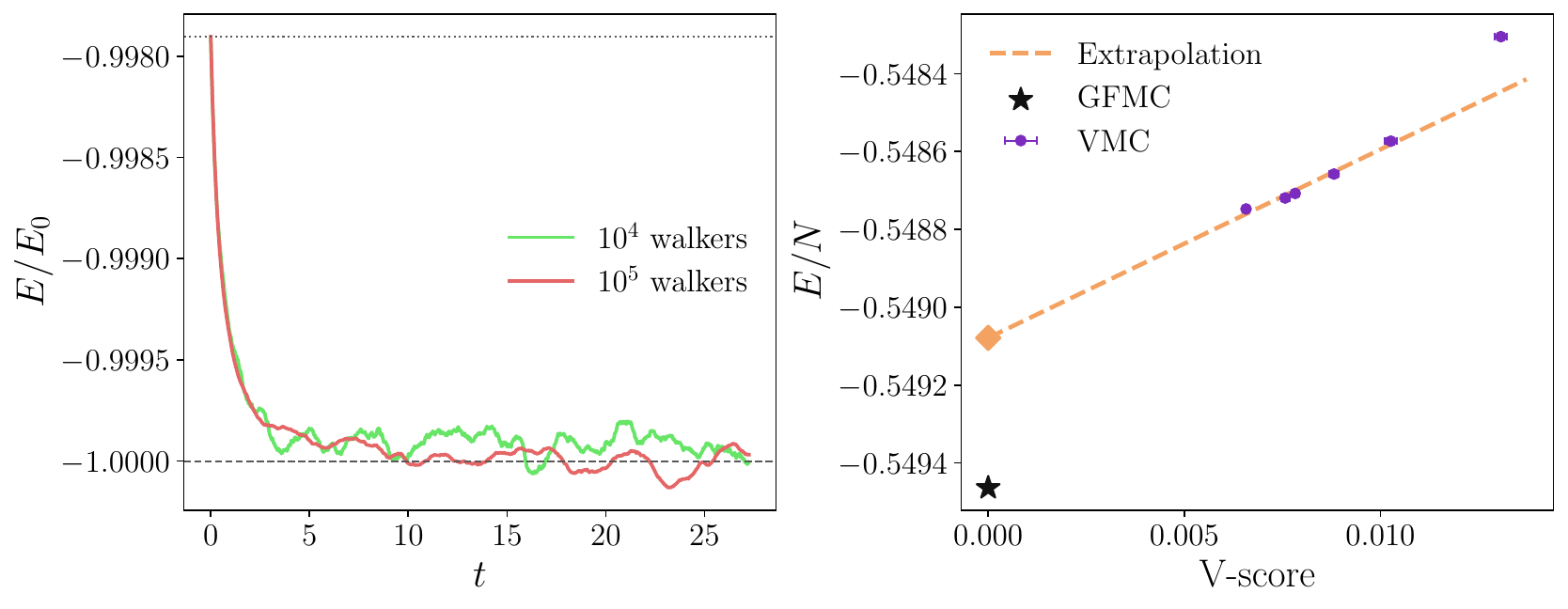}
  \caption{\textbf{Error estimation methods.} Left: GFMC time evolution at $N=10^4$, $h_\text{x}=0.45$ for both $n_w=$\green{$10^{4}$} and \red{$10^{5}$}.
  Right: Zero-noise extrapolation for $N=10^5$, $h_\text{x}=0.45$. The black star represents the energy obtained with GFMC. }
  \label{fig:error_estimation}
\end{figure}

As important as the equation of state itself would be the error
on this equation of state as its analysis contains insights on its
origin, e.g., deviation to adiabaticity close to gap closure versus presence of noise or decoherence in the experiment.
The lower panels of Fig.~\ref{fig:eq_of_state} show this error for our VMC ansatz. To compute the error $\delta E$, we have used different complementary methods that we have found to be equivalent. The first is to use GFMC as the reference energy
(the gf-error, see the left panel of Fig.~\ref{fig:error_estimation} for a typical GFMC trace in imaginary time). The second method is to use the variance of the VMC ansatz as a proxy of the error; this method is presented in Appendix \ref{app:vscore}.
For problems too large for us to run the GFMC simulations, we used a third (consistent) method to compute the error: zero-noise extrapolation \cite{kashima2001} (the zne-error).
The zne-error method is illustrated in the right panel of Fig.~\ref{fig:error_estimation}: one simply extrapolates the energy versus variance curve obtained during the minimization to zero variance. Notice the scale of the extrapolation which only affects the third digit in the energy.
 Note that the experimental precision is not sufficient to ensure that the data lie strictly within the linear regime of zero-noise extrapolation, resulting in a lower accuracy.

These results for $\delta E/E_0$ are relevant for two reasons:
First, they provide the order of magnitude of what can be obtained with a very simple ansatz, with a very low computational cost. We find that the relative error is always smaller than $10^{-2}$ in VMC (and $10^{-3}$ or lower for the corresponding GFMC). Second, we find that the maximum error actually matches the position of the gap-closing (where the paramagnetic to anti-ferromagnetic transition happens) which is precisely the position where a QA would face Landau-Zener transitions. To compute the position of the gap-closing, we have used a GFMC-based method which we call Gap-FMC, and which is, to the best of our knowledge, original. The Gap-FMC method is explained in \cref{app:gap}.

A key feature of our ansatz is that it is computationally
very light, allowing us to scale the simulations up to
$N=100,000,000$. The corresponding energy versus size is shown on the top panel of \cref{fig:scaling_error}. On a 64 cores node with 256GB of Memory, we could run 25 walkers in parallel in roughly 3 hours for the $N=10^8$ point. Note that a single walker is sufficient at this size, as indicated by the cumulative distribution of the local energy shown in the inset (all our $200$ samples lie within $10^{-5}$ of each other).

The lower panel of Fig.~\ref{fig:scaling_error} shows the error of our simulations
versus $N$ for our variational ansatz (circles when we used the gf-error and squares when we used the zne-error).
 Importantly, the error does \emph{not} deteriorate with the size $N$. Indeed, as the system gets bigger, the relative contribution of the surface to the energy gets smaller and the quality of our ansatz (designed for a system invariant by translation) improves.  \cref{fig:scaling_error} also shows the error of the GFMC simulation for $n_w=10^{4}$. The error of the GFMC run was estimated by looking at the fluctuations of $E(\tau)$ at large $\tau$ as well as the difference to  a more precise $n_w\ge 10^{5}$ reference calculation (see the left panel of \cref{fig:error_estimation}).
For the largest size, $N=317\times 317$ indicated by \textcolor{mystarblue}{$\bigstar$},
we could run the GFMC simulation only for $n_w=10^4$ walkers so the corresponding error estimate is essentially an extrapolation from smaller $N$ values; we don't have a reliable estimate of the error for this point.
Also shown is a highly accurate calculation at $N=100$, $h_\text{x}=-0.45$ using a more sophisticated ansatz.
This is a NQS ansatz built with perceptrains using the technique described in \cite{srdinsek_2025}.
Last, the orange square shows our estimate of where the current experimental annealers are, based on the noise levels observed in the cited literature and private discussions. This is merely an estimate since we are not aware of reported experimental data on the equation of state.

In order to reach such a large size $N=10^8$ with a reasonable computational budget, on a single core, we have performed several optimizations.
The first one is to truncate the range of the interaction to a radius $R$. Since the interaction decays as $1/r^6$, this gives rise to an error that decays rapidly as $O(1/R^4)$. We choose
$R=8.7$ for $N \leq 10^4$ and $R=4.1$ for $N > 10^4$ points in order to ensure that the range truncation error remains below the reported  error, see \cref{app:range} for more details. This approximation induces a reduction of the memory footprint by a factor $\propto R^2 /N$, and a computational gain of a factor $\propto R^2 \log (N)/N$. The range truncation error can also be interpreted as a limitation of the hardware: for a given QUBO problem (fixed, say, by an industrial application), it is unlikely that the matrix $J_{ij}$ had the $1/r^6$ form provided by the Rydberg atom platform. The solution to this problem is usually to embed the QUBO problem into a larger system; however, in that setting the inability to switch off the couplings at large distance will limit the accuracy at which one can approach the prescribed matrix $J_{ij}$.
 A second optimization relies on the fact that the statistical error of the calculation of the energy is self-averaging (see \cref{app:vscore}) so that the number of walkers can be decreased as $n_w\sim 1/N$ at constant statistical accuracy. We have typically used $n_w = 10^4$  for the smaller sizes but $n_w=1$ is sufficient to resolve the energy of the largest VMC run.
 GFMC, however, does not benefit from the self-averaging and all GFMC runs were performed with $n_w=10^4$. 
 Last, the effective temperature optimization has been done up to $N=1600$. Above this value, we have kept the parameters obtained at $N=1600$ which appeared to have converged (a small accuracy gain can probably be obtained at the cost of relaxing this constraint).

\begin{figure}[htbp]
  \centering
  \includegraphics[width=\linewidth]{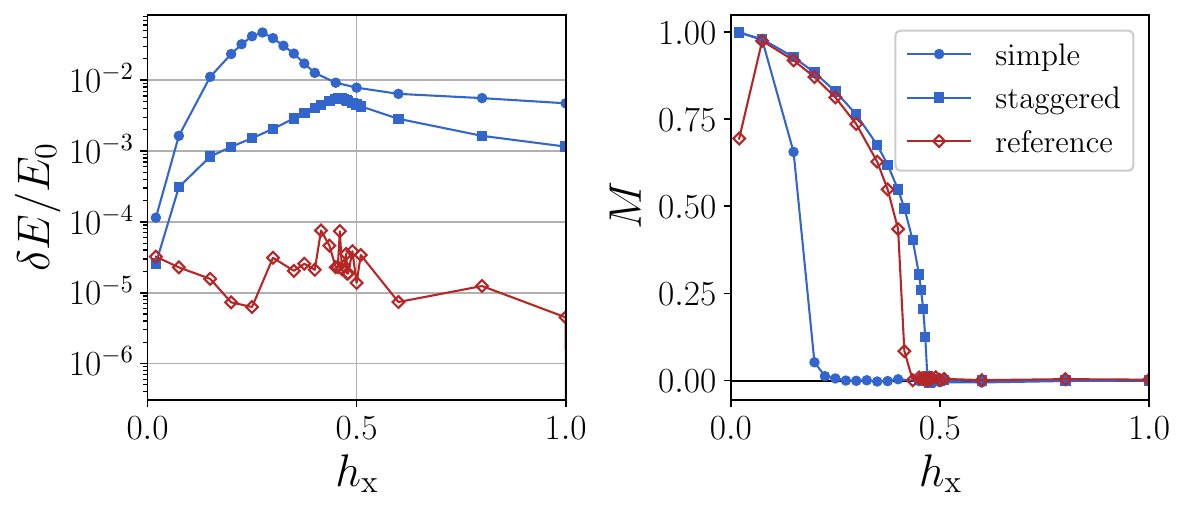}
  \caption{\textbf{Role of the parameters beyond simple ansatz.}
  Relative error (left) and staggered magnetization (right) for the square lattice $N=100$ and $h_\text{z}=-1$. The three curves correspond to
  the simple thermal ansatz \cref{eq:simple}, the staggered thermal ansatz \cref{eq:ansatz_stag} and a highly accurate Perceptrain calculation using the algorithm described in Ref.~\cite{srdinsek_2025}.
 }
  \label{fig:err_vs_obs}
\end{figure}

To put the accuracy values into perspective, we end this section
by a discussion of the staggered magnetization $M$, i.e., not the
energy itself but its first derivative. A typical feature observed in variational calculations is that it is relatively easy to arrive at a decent accuracy for the equation of state, but small energy scales can be associated with important changes of some observable. For QUBO, this is probably a good thing. It means that it is relatively easy to reach a reasonable accuracy. Conversely, the quality of a QUBO solver is often hidden in the difficulty to access the last digits which may be associated to large changes in the solution. \cref{fig:err_vs_obs} shows the simple ansatz error, the staggered ansatz error and a reference perceptrain calculation error versus $h_\text{x}$ (left panel). The right panel shows the staggered magnetization $M$. The reference has been obtained using the highly accurate technique discussed in Ref.~\cite{srdinsek_2025}. We observe that an accuracy below $10^{-2}$ -- which we obtain with the staggered ansatz but not with the simple ansatz \cref{eq:simple} -- is necessary to reproduce qualitatively the staggered magnetization. Interestingly, the staggered ansatz predicts a phase transition at a slightly higher $h_\text{x}$ than the reference. We attribute this observation to the fact that the
ansatz is translationally invariant, hence insensitive to the presence of the large surface, which slightly shifts the position of the transition.
Investigating in more details the accuracy of the ansatz, our variational wavefunction was able to detect that both $h_\text{x}=0$ ground states of the square lattice ($h_\text{z}\approx 2.33$) and triangular lattice ($h_\text{z}=1.5$) had a corner defect (see \cref{app:deffect} for more details). It is unclear if the experiments \cite{ebadi2021,scholl2021} observed this corner defect or not. Since it is associated with a $\sim 1\%$ error, it is a good indication of the precision of the experiments.  For the triangular lattice, the thermal ansatz was able to detect a staggered magnetization hysteresis at a transition believed to be first order \cite{janke1997,guo2023} (see \cref{app:firstorder}).

\section{Benchmark of a QUBO problem}\label{sec:qubo}

\begin{figure}[htbp]
  \centering
  \includegraphics[width=\linewidth]{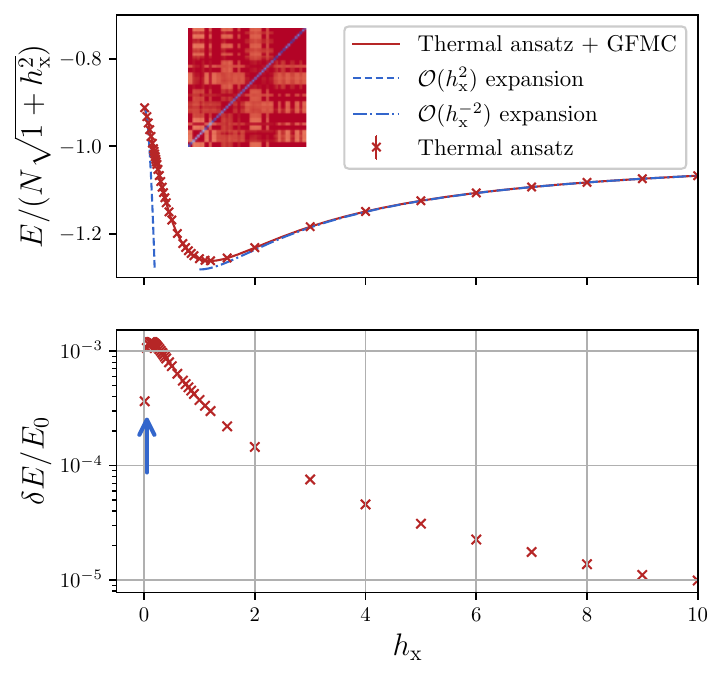}
  \caption{\textbf{Equation of state of a fallen angel instance.} Top: Energy as a function of transverse field computed with the simple thermal ansatz (crosses) and simple thermal ansatz plus GFMC (line) for an instance of
 \cite{leclerc2022financial}. Inset: colormap of the QUBO matrix $J_{ij}$ (red: positive, blue: negative). Bottom: relative error of the simple thermal ansatz (using the GFMC computation as a reference).
All energies have been rescaled by a factor $1000N$ and fields by a factor $1000$ to keep the typical scales of order unity.}
  \label{fig:eq_of_state_qubo}
\end{figure}

We end this article with $N=50$ matrices $J_{ij}$ that originate from a practical QUBO use case. Five different such matrices were provided to us by the authors of Ref.~\cite{leclerc2022financial}; they correspond to
a financial risk management problem known as "fallen-angels forecasting" for loans attribution. Here, we used the original matrix $J_{ij}$ \emph{before} it was (approximately) mapped onto a Rydberg atom problem, since the latter was not recorded by the authors. The color map of one of the matrices is shown in the inset of \cref{fig:eq_of_state_qubo}; it consists of relatively large, apparently disordered, anti-ferromagnetic cross-like structures (red) frustrated by a large negative diagonal (blue).

The equation of state for this instance is shown in \cref{fig:eq_of_state_qubo}. It is obtained with the simple ansatz \cref{eq:simple}, yielding a relative error of at most $\approx 10^{-3}$.
In contrast to the previous simulations of physical systems, the maximum of the error appears to be near $h_\text{x}=0$; our GFMC calculation confirms that the gap seems to be indeed vanishing there; see \cref{app:gap}.
We observe that the simple ansatz is rather accurate for these instances.
As previously, it could be further improved by enlarging the parameter space, for instance using an ansatz of the form $E^\text{eff}(\boldsymbol{n}) = \beta^\text{eff} F(\boldsymbol{n}) + \sum_i \beta^{\text{eff}}_i n_i$.

\begin{figure}[htbp]
  \centering
  \includegraphics[width=\linewidth]{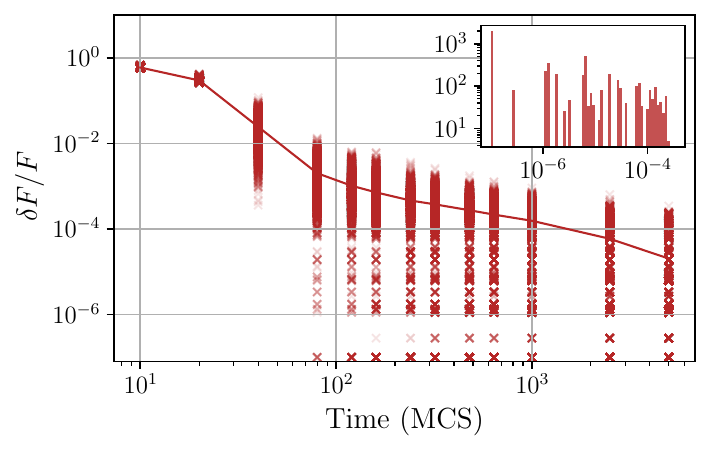}
  \caption{\textbf{Simulated Quantum Annealing.} Relative score $\delta F /F$ of the simple thermal ansatz SQA as a function of the time expressed in the number of Monte Carlo steps (1 MCS took us $12 \mu s$ on one core).
  The results (lines) are averaged over the five different QUBO instances and $n_w=1000$ per instance. The crosses show the different
  values of the error during the annealing.
 Shots reaching the benchmark solution ($\delta F / F =0$) are artificially placed at $\delta F / F=10^{-7}$ for clarity.
 Inset: histogram of $\delta F / F$ after $5.10^{3}$ MCS, at the end of the annealing process.}
  \label{fig:qubo_err}
\end{figure}

Last, we actually perform a simulated quantum annealing protocol to
solve the corresponding QUBO problem. Denoting $\bar{\boldsymbol{n}}$ the actual solution of the problem (which we have found using different long runs as well as independent thermal annealing simulations), we measure the accuracy of the solution as
\begin{equation}
  \frac{\delta F}{F} = \frac{F (\boldsymbol{n})-F (\bar{\boldsymbol{n}})}{|F (\bar{\boldsymbol{n}})|}.
\end{equation}
The results are shown in \cref{fig:qubo_err}.
As expected, the results strongly depend on the number of Monte-Carlo steps (MCS, i.e., how fast we perform the annealing from $h_\text{x}=1$ to $h_\text{x} = 0.01$) as well as the number of walkers used. At 12 $\mu$s/MCS/core, the five instances could be solved almost instantly on a single core (typically in a few ms). This is typically faster by a factor $10^6$ than the one thousand repetitions at one Hertz used to reach the $1\%$ threshold in the experiments of Ref.~\cite{leclerc2022financial}. More interestingly, we find that error $\delta E/E_0$ is much more instructive than $\delta F/F$: the former is an intrinsic measure of what the hardware/ansatz can do while the latter can be made arbitrarily small by increasing the number of shots.

%% file: sections/appendix_beta.tex
\section{Effective temperatures and symmetries}\label{app:beta}

\begin{figure}[htbp]
  \centering
  \includegraphics[width=\linewidth]{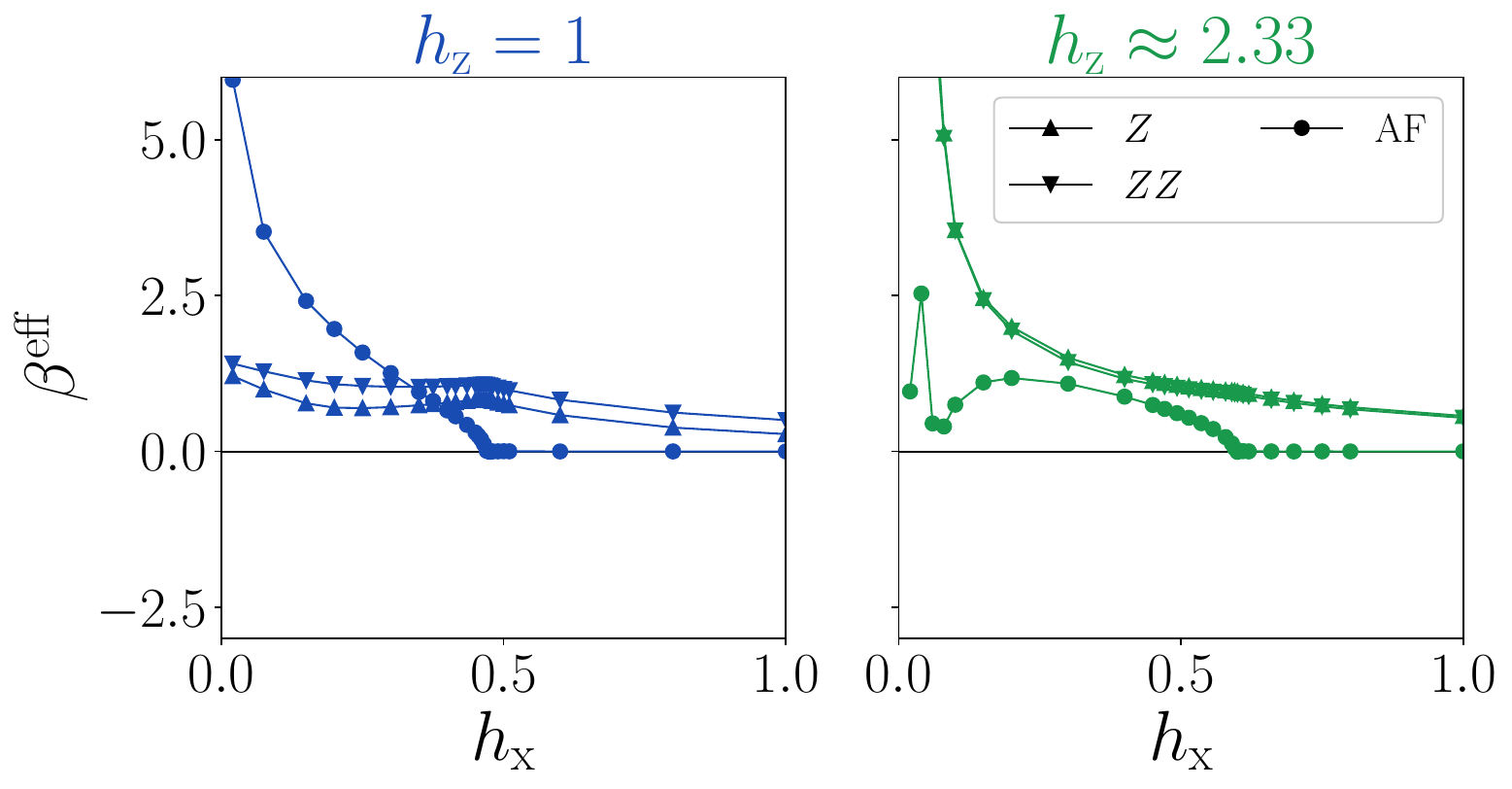}
  \caption{\textbf{Optimum effective inverse temperatures $\beta_\text{z}$,
  $\beta_\text{zz}$ and $\beta_\text{AF}$ versus transverse field $h_\text{x}$.} These temperatures were obtained for the staggered thermal ansatz \cref{eq:ansatz_stag} for the square lattice in the simulations of \cref{fig:eq_of_state}. Left panel: $h_\text{z}= 1$. Right panel: $h_\text{z}\approx 2.33$. See text for the definitions.
  }
  \label{fig:betas}
\end{figure}

\cref{fig:betas} shows the optimum effective temperatures that we have found for the square lattice, at two values of $h_\text{z}$. We have used the following definitions: $\beta_\text{zz} \equiv \beta^\text{eff}$ (for the
interaction terms), $\beta_\text{z} \equiv (\beta_0 + \beta_1)/(2h_\text{z}) +\beta^\text{eff}$ (for the zeeman field) and $\beta_\text{AF} \equiv (\beta_0 -\beta_1)/2$ (for the staggering field).

Two interesting remarks can be made from these data.
First, we observe a coincidence between the appearance of a
non-zero staggered field $\beta_\text{AF}$ and the presence of an actual non-zero staggered magnetization $M$ in the problem.
 This is, in our view, an advantage of
having a physically motivated variational ansatz; it is highly interpretable.
Second, the one-body inverse temperature $\beta_\text{z}$ and the two-body interaction inverse temperature $\beta_\text{zz}$ have the same value
for $h_\text{z}\approx 2.33$ (right) but not for $h_\text{z}=1$ (left).

This second fact can be explained by a symmetry argument. In the hole basis $\bar{\mathbf{n}}$ defined as $\bar{n}_i = 1 - n_i$, the transverse field Ising Hamiltonian \cref{eq:ising} remains invariant with, up to an irrelevant constant term,
\begin{align}
\bar{J}_{i\ne j} &= J_{ij} \\
\bar{J}_{ii} &= -J_{ii} -2\sum_{j\ne i} J_{ij}.
\end{align}
In the case of a square lattice of Rydberg atoms, and in the thermodynamic limit, the transformation corresponds to a symmetry $h_\text{z}\rightarrow (2h_\text{z}^\text{c}-h_\text{z})$ around the magnetic field,
\begin{equation}
  h_\text{z}^\text{c} \equiv \sum_{i\neq0} \frac{1}{2|\vec{r}_i|^6} \approx 2.33.
\end{equation}
Interestingly, our variational staggered ansatz automatically inherits the same symmetry
\begin{align}
  \psi_{\boldsymbol{n}}(h_\text{z}) =
  \psi_{\boldsymbol{1}-\boldsymbol{n}}(2h_\text{z}^\text{c}-h_\text{z}).
\end{align}
By construction, the effective temperatures satisfy
\begin{align}
  &\beta_\text{zz}(h_\text{z}) = \beta_\text{zz}(2h_\text{z}^\text{c}-h_\text{z}) \\
  &h_\text{z}\beta_\text{z}(h_\text{z}) = -(2h_\text{z}^\text{c}-h_\text{z}) \beta_\text{z}(2h_\text{z}^\text{c}-h_\text{z}) +\beta_\text{zz}(2h_\text{z}^\text{c}-h_\text{z})2h_\text{z}^\text{c}\\
  &\beta_{\text{AF}}(h_\text{z}) = -\beta_{\text{AF}}(2h_\text{z}^\text{c}-h_\text{z}).
\end{align}
It follows from those relations that $h_\text{z} =h_\text{z}^\text{c}$ implies $\beta_\text{z}(h_\text{z}) = \beta_\text{zz}(h_\text{z})$ as observed in our numerics.
More generally, it is clear that the symmetries of the studied Hamiltonian can be directly propagated to the thermal ansatz (that shares a very similar structure) thereby reducing the number of parameters and the learning computational cost.
Reversely, in this article we only considered a translationally invariant ansatz while the system itself doesn't have this perfect symmetry.
 This makes it clear that relaxing this constraint should allow lower error in the ground state (as we observed when increasing the size of the system). Finally, the staggered field $\beta_{\text{AF}}$ of the staggered ansatz can be seen as a way to break the Hamiltonian symmetries or equivalently to enforce the symmetries of the anti-ferromagnetic phases or more generally of any desired phases.

%% file: sections/appendix_edge.tex
\section{Corner defects}\label{app:deffect}
An important constraint of quantum hardware with respect to numerical simulations is the presence of large finite size effects. A typical way to alleviate those is to use periodic boundary conditions, which are typically not available in an experimental platform. For $N=100$ Rydberg atoms on a square lattice, more than one third ($36$) of the atoms lie on the edges.
It has been predicted that a separate transition could take place on the boundary before the one taking place in the bulk \cite{kalinowski_2022}.

This point can be understood already in the classical limit
$h_\text{x}=0$: starting from a configuration $\mathbf{n}$ where the site $i$ is unoccupied (say the perfect anti-ferromagnetic configuration to be concrete), the energy $\Delta E$ to fill this state reads
\begin{equation}
  \Delta E = \sum_{j\ne i} J_{ij}n_j + h_\text{z}.
\end{equation}
It follows that the field needed to fill the state ($\Delta E=0$) is significantly smaller in the corners of the system than in the edges and in the edges than in the bulk (the sum contains less terms). For the anti-ferromagnetic square lattice, this criterion gives $h_\text{z} \gtrsim 2.02$ for the "corner transition". Our numerical experiment agrees with this estimate in various ways. Indeed, this defect is not observed in our numerical result for $h_\text{z}=1$, but it does emerge for $h_\text{z}\approx 2.33$ on the square lattice and for $h_\text{z}=1.5$ on the triangular lattice (where the corner defect requires $h_\text{z} \gtrsim 1.04$) as $h_\text{x}$ goes to zero.

\begin{figure}[htbp]
  \centering
  \includegraphics[width=\linewidth]{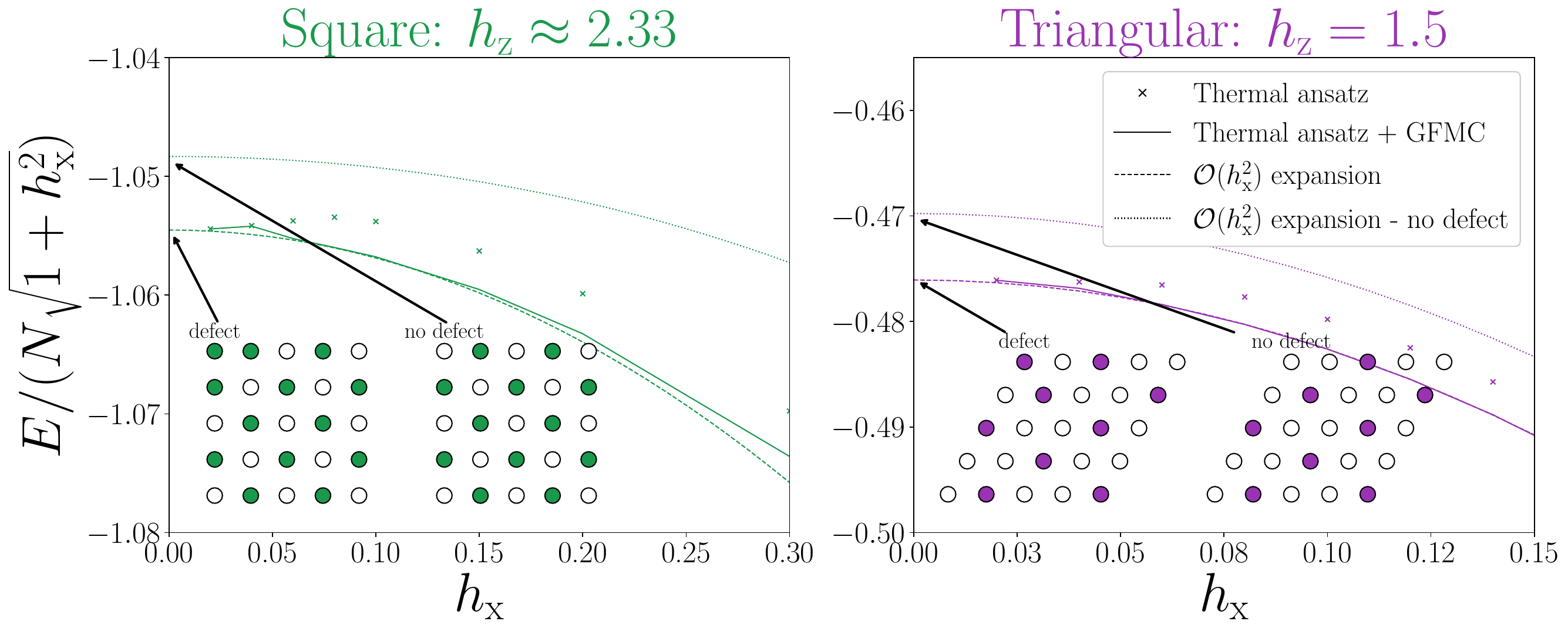}
  \caption{\textbf{Effect of corner defects on the equation of state.}
  Zoom of the equation of state $E(h_\text{x})$ at small field.
  The energy matches the perturbation theory only when the corner defect
  is properly accounted for.
  Left panel: square lattice for $h_\text{z}\approx2.33$ and $N=100$ atoms. Right panel: triangular lattice for $h_\text{z}=1.5$ and $N=147$ atoms.
  Inset: snapshot of the $h_\text{x}=0$ state with and without a defect at the corner.}
  \label{fig:defect}
\end{figure}

Our calculated equation of state matches the prediction with a defect as shown in \cref{fig:defect}. The corner defect can also be seen directly by looking at snapshots of the configurations at small $h_\text{x}$ or by computing the staggered magnetization $M$, which does not reach the pure anti-ferromagnetic limit $M=1$. For the square lattice, there are two corner defects opposed along the diagonal for even $N$ but none for odd values of $N$. In the case of the triangular lattice with $N=147$ atoms (see \cref{fig:phase_diag}), we also observe two corner defects for $h_\text{z}=1.5$.
Finally, our gap computation at small $h_\text{x}$ in \cref{fig:gap} agrees with the presence of a corner defect at $h_\text{z}\approx 2.33$ and its absence at $h_\text{z}=1$.

%% file: sections/appendix_perturbation.tex
\section{Perturbation theory}\label{app:perturbation}
The Hamiltonian $\hat H$ has trivial ground states both at zero
$h_\text{x}=0$ and infinite $h_\text{x}\gg \max_{ij} |J_{ij}|$ fields.
In this appendix, we provide the (straightforward) expressions (shown in the main text) for the second order perturbation theories around these two limits.

\subsection{Weak field expansion ($h_\text{x} \ll J_{ij}$)}
Let  $\ket{\tilde{\boldsymbol{n}}}$ be the ground state of $\hat H_\text{z}$. It is typically the antiferromagnetic state
(possibly with a corner defect, see \cref{fig:defect}).
Perturbation theory in $h_\text{x}$ for
 $\hat H = \hat H_\text{z} - h_\text{x} \hat H_\text{x}$ gives
\begin{equation}
    E_0 = F(\tilde{\boldsymbol{n}}) - h_\text{x}^2 \sum_i \frac{1}{\delta_i} +\mathcal{O}(h_\text{x}^3),
\end{equation}
with $\delta_i$ defined as
\begin{equation}
    \delta_i =  J_{ii} + 2(1-2\tilde{n}_i) \sum_{j} J_{ij}\tilde{n}_j.
\end{equation}

\subsection{Strong field expansion ($h_\text{x} \gg J_{ij}$)}
At large field, the reference state is the ground state of $\hat H_\text{x}$, i.e., $ |+\rangle^{\otimes N}$. Apply perturbation theory to $\hat H/h_\text{x} = \hat H_\text{x} - \frac{1}{h_\text{x}} \hat H_\text{z}$.
 The perturbation $\hat H_\text{z}$ contains terms acting on either one site
 $\hat n_i$ or two sites $\hat n_i \hat n_j$.
 This results in the ground state energy
\begin{align}
E_0 &= -h_\text{x} N + \frac{1}{2}\sum_{i\leq j}J_{ij} \\
&-\frac{1}{h_\text{x}} \left( \frac{1}{8}\sum_i \left(\sum_j J_{ij}\right)^2
+\frac{1}{16}\sum_{i<j} J_{ij}^2 \right) +\mathcal{O}(h_\text{x}^{-2}).\nn
\end{align}

%% file: sections/appendix_gap.tex
\section{Gap estimation with Gap-FMC}\label{app:gap}
\begin{figure}[htbp]
  \centering
  \includegraphics[width=\linewidth]{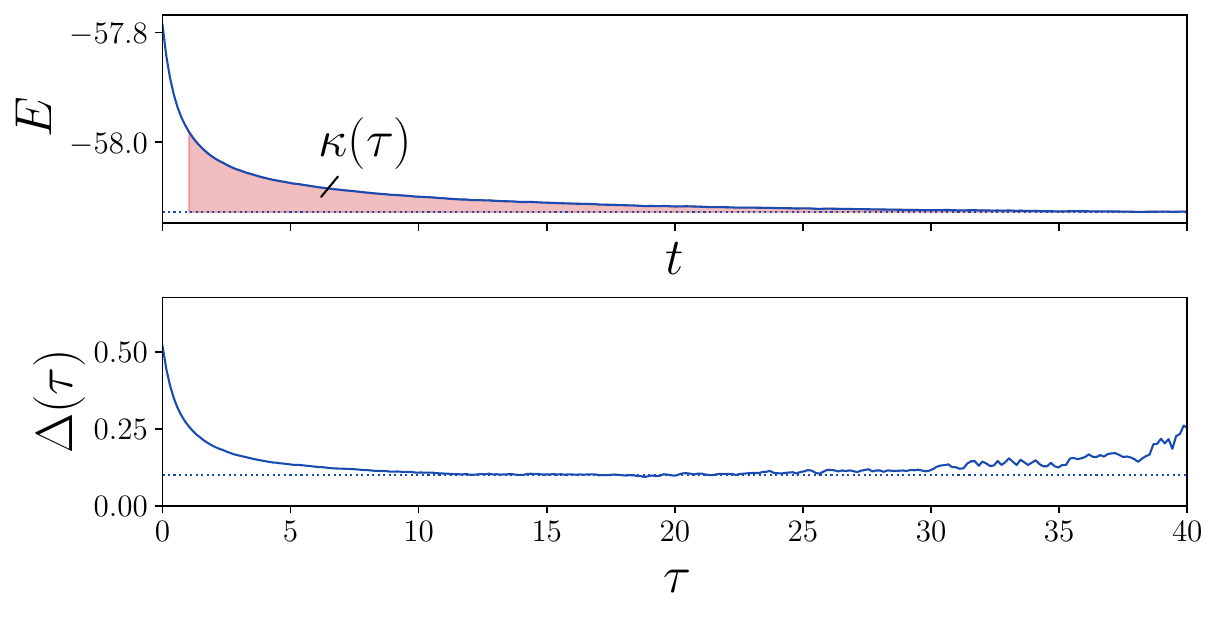}
  \caption{\textbf{The GapFMC method for gap estimation.}  Top panel: GFMC trace of energy $E$ as a function of imaginary time $t$ for $h_\text{x}=0.45$ and $h_\text{z}=1$.
  We define $\kappa (\tau)$ as the area below the curve between $\tau$ and $+\infty$.
  Bottom: the time dependent gap estimate $\Delta (\tau)$ converges to the actual gap $\Delta$ (estimated by the horizontal line) and eventually becomes noisy. This computation was done with a large number of $n_w= 10^7$ walkers for a $10$ by $10$ square grid. }
  \label{fig:method_gap}
\end{figure}

In this appendix, we introduce a GFMC based method to extract the gap of the Hamiltonian, hereafter called the GapFMC method. Let $\hat H$ be a Hamiltonian with eigenenergies $E_0$, $E_1$, $E_2$, \ldots, and corresponding eigenstates $\ket{0}$,  $\ket{1}$,  $\ket{2}$, \ldots. We want to estimate its
gap $\Delta = E_1-E_0$. We further define $\Delta_2 = E_2-E_1$.
Let
\begin{equation}
\ket{\psi(\tau)}= e^{-\tau \hat H}\ket{\psi_\text{V}}.
\end{equation}
GFMC provides a direct way to calculate the energy versus imaginary time
curve
\begin{equation}
  E(\tau)= \frac{\braket{\psi_\text{V}|\hat H|\psi(\tau)}}{\braket{\psi_\text{V}|\psi(\tau)}}.
\end{equation}
An example is shown in the top panel of \cref{fig:method_gap}.
A direct generalization of the result of \cite{mora2007} is that the overlap
\begin{equation}
\Omega(\tau) \equiv |\braket{0|\psi(\tau)}|^2
\end{equation}
can be obtained from the area under the curve $E(\tau)$. More precisely,
\begin{equation}
  \Omega(\tau)=e^{-\kappa (\tau)}, \kappa(\tau)=\int_{\tau}^{\infty} dt \ [E(t)-E_0].
\end{equation}
Now, for large time, and up to a global normalization, we have,
\begin{equation}
\ket{\psi(\tau)} = \sqrt{\Omega(\tau)}\ket{0} + \sqrt{1-\Omega(\tau)}
\ket{1} + O(e^{-\Delta_2\tau}).
\end{equation}
Now let us define
\begin{equation}\label{eq:gap_est}
  \Delta(\tau) \equiv \frac{E(\tau)-E_0}{1-\Omega(\tau)}.
\end{equation}
It follows from the above definitions that $\Delta(\tau)$ converges towards $\Delta$
\begin{equation}
 \Delta(\tau)  = \Delta + O(e^{-\Delta_2 \tau}),
\end{equation}
provided the overlap does not vanish. Also, extending again \cite{mora2007}, the convergence takes place from above:
\begin{equation}
\Delta(\tau) \ge\Delta
\end{equation}
 Note that a similar approach based on path integral quantum Monte Carlo was independently developed in parallel in~\cite{brodoloni2026}. The lower panel of \cref{fig:method_gap} shows an example of the convergence of $\Delta(\tau)$.
We note that the method is quite demanding computationally.
Indeed, for large $\tau$, $\Omega(\tau)$ gets close to one (vanishing area $\kappa(\tau)$) while $E(\tau)\rightarrow E_0$. Hence $\Delta(\tau)$ is the ratio of two quantities that vanish asymptotically -- not an ideal situation. Nevertheless, for a finite value of $\tau$ and using very precise
simulations (large number of walkers), one can actually compute the gap which provides valuable insights.

\begin{figure}[htbp]
  \centering
  \includegraphics[width=\linewidth]{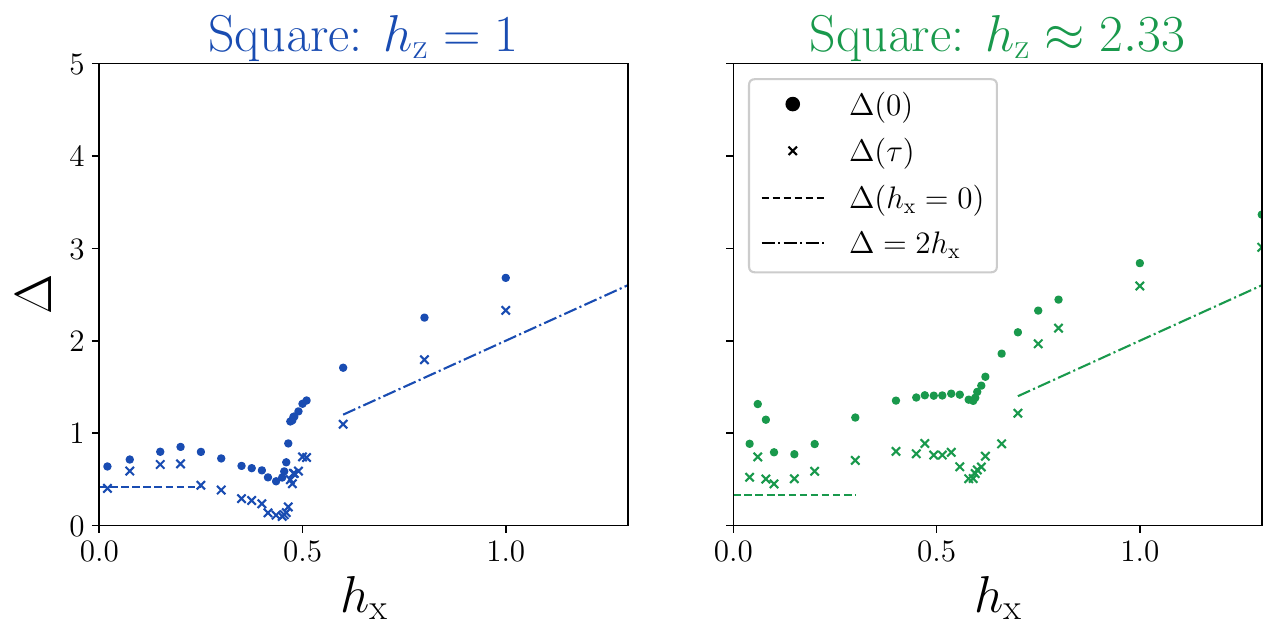}
  \caption{\textbf{Closing of the gap for the square lattice of Rydberg atoms.} Gap estimation $\Delta$ as a function of the transverse field $h_\text{x}$. Left panel:  $h_\text{z} = 1$. Right panel: $h_\text{z}\approx 2.33$. The calculations were performed with $10^7$ walkers for the
  $N= 10\times 10$ square system.}
  \label{fig:gap}
\end{figure}

We apply this method to measure the gap for the two annealing paths on the square lattice; the results are shown in \cref{fig:gap}.
We observe a gap closing at the phase transition between the PM and the AF phases. Our gap estimate also has the correct asymptotic limits for large (dashed dot lines) and vanishing (dashed line) values of $h_\text{x}$. We emphasize that the small $h_\text{x}$ limit gap at $h_\text{z} \approx 2.33$
wouldn't be correct if we didn't take the corner defect (see \cref{app:deffect}) into account in the computation.

\begin{figure}[htbp]
  \centering
  \includegraphics[width=\linewidth]{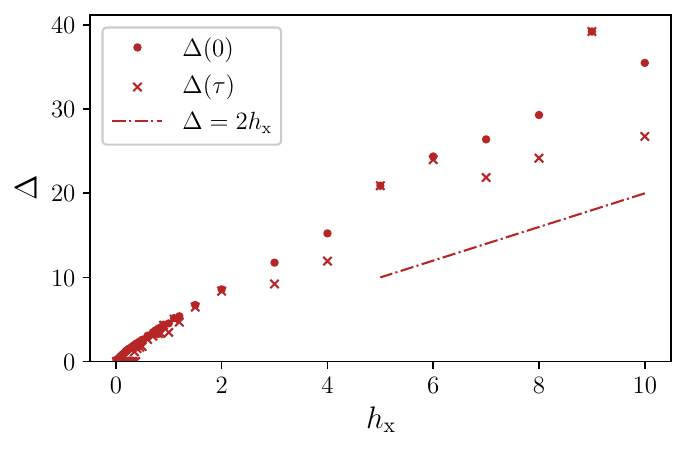}
  \caption{\textbf{Closing of the gap for the fallen angel problem.} Gap estimation $\Delta$ as a function of the transverse field $h_\text{x}$. The computation was performed with $5.10^7$ walkers for one QUBO
  instance of \cite{leclerc2022financial}, $N=50$. The transverse field and the gap are rescaled by a factor $1000$.}
  \label{fig:gap_qubo}
\end{figure}

\cref{fig:gap_qubo} shows our gap estimate for the fallen angel QUBO problem of \cite{leclerc2022financial}. As advertised, we do observe a gap closing
that takes place at vanishing (or at least very small) transverse field $h_\text{x}$. All of the 5 individual QUBO instances we had access to had similar gap profiles.

%% file: sections/appendix_vscore.tex
\section{Variance scaling and V-score}\label{app:vscore}

\begin{figure}[htbp]
  \centering
  \includegraphics[width=\linewidth]{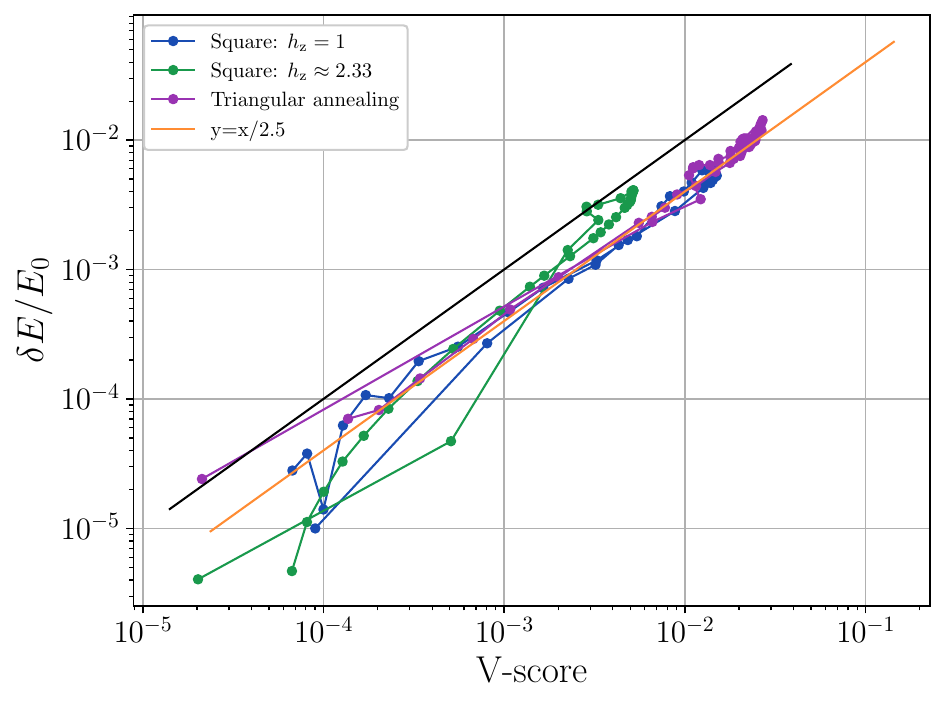}
  \caption{\textbf{Relative error versus V-score.} $\delta E/E_0$ as a function of the V-score $N \sigma^2/E^2$ along the different annealing paths of \cref{fig:eq_of_state} (square lattice $N=100$ $h_\text{z}=1$,
   $h_\text{z}\approx2.33$ and triangular lattice $h_\text{z}=1.5$).
   We observe that the V-score roughly corresponds to the number of correct digits in the energy ($\delta E/E_0$ is actually $2.5$ times smaller than V-score).}
  \label{fig:vscore_prop}
\end{figure}

In this appendix, we briefly discuss the variance of the energy $\sigma^2$,
\begin{equation}
\sigma^2 \equiv \frac{ \braket{\psi_\text{V}|\hat{H}^2|\psi_\text{V}}}{\braket{\psi_\text{V}|\psi_\text{V}}} - E(h_\text{x})^2
\end{equation}
and its scaling with $N$. It is expected that both the energy \emph{and}
its variance are extensive quantities, i.e., $E= e N$ and
$\sigma^2 \propto N$.

To understand where this scaling comes from, let us consider a local Hamiltonian
\begin{equation}
  \hat H = \sum_{i} \hat h_i,
\end{equation}
where $\hat h_i$ acts on a site $i$ or a small group of $O(1)$ sites around it. One finds that
\begin{equation}
\sigma^2 = \sum_{ij} C_{ij}
\end{equation}
where the energy density correlation function $C_{ij}$ is defined by
\begin{equation}
C_{ij} \equiv \braket{\hat h_i \hat h_j} - \braket{\hat h_i} \braket{ \hat h_j}.
\end{equation}
For a translationally invariant gapped system, correlation functions
are expected to decay exponentially as $C_{ij} \sim e^{-|\vec r_i -\vec r_j|/\xi}$ from which it follows that the variance is extensive
\begin{equation}
\sigma^2 = N \sum_{j} C_{0j},
\end{equation}
which is almost universally observed in VMC calculations (the argument actually survives a weaker decay of the correlations).

This property has two important consequences.
First, the statistical noise in variational Monte Carlo scales as
\begin{equation}
  \frac{\delta E_\text{stat}}{E_0} = \frac{\sqrt{\sigma^2/n_w}}{E_0} = \frac{1}{e_0}\sqrt{\frac{\sum_{j} C_{0j}}{n_w N}},
\end{equation}
i.e., the relative statistical accuracy improves with $N$ for a fixed number $n_w$ of walkers. In other words, there is a certain level of self-averaging
in the statistical error. This point can be used to reduce the number of walkers as $n_w \propto 1/N$ for large-system simulations.
Second, the variance can be used as a good proxy of the error, remembering that it vanishes when $\ket{\psi_\text{V}}$ approaches the actual ground state of the system. To account for the scalings, one considers the so-called V-score,
an intensive quantity defined as $\text{V-score}=\frac{N \sigma^2}{E^2}$.
It has been observed repeatedly that the V-score is proportional to the relative error \cite{kashima2001, wu2024, srdinsek_2025},
$\delta E/E_0 \propto \text{V-score}$.
\cref{fig:vscore_prop} shows the error $\delta E/E_0$ plotted as a function of $\text{V-score}$ for various data points studied in this article.
We find indeed a very good proportionality relation so that the
V-score can be used as a (rough) proxy estimator of the error.

%% file: sections/appendix_range.tex
\section{Large scale simulation and range truncation}\label{app:range}
In this appendix, we provide some details about the optimizations used to perform the largest simulations of this article.
Let us first consider the single-spin Hastings-Metropolis to sample our thermal ansatz distribution
\begin{equation}
  P(\boldsymbol n) \approx e^{- E^\text{eff}(\boldsymbol n)}.
\end{equation}
A site $i$ flip is randomly proposed and accepted with probability
\begin{equation}
  \min\{1,e^{-\Delta_i E^\text{eff}}\},
\end{equation}
where $\Delta_i E^\text{eff}$ is the change of effective energy under this
single flip. Since the energy follows \cref{eq:couplings} calculating $E^\text{eff}(\boldsymbol{n})$ of a configuration takes $N^2$ operations.
However, the sampling procedure only requires to compute the energy difference. The latter can be written
\begin{equation}
  \Delta_i E^\text{eff}
  =
  \bar{n}_i\left[
  J^\text{eff}_{ii}
  +2\sum_{j\ne i}J^\text{eff}_{ij}n_j
  \right],
  \label{eq:delta_metropolis}
\end{equation}
where $\bar{n}_i \equiv (1-2n_i)$, the calculation of which costs only $N$ operations. Since the interaction between Rydberg atoms decays as $1/r^6$, an approximate energy can be obtained up to a correction $O(1/R^4)$ by
truncating the interaction beyond a certain range $R$ (more on that below). Such an approximation
reduces the computation cost of $\Delta_i E^\text{eff}$ from $N$ to $O(R^2)$ without significantly altering the results.
Overall, to produce a fully thermalized sample, one needs to flip each spin at least once, which requires $O(N)$ random moves for an overall computational cost that scales as,
\begin{equation}\label{eq:discrete_scaling}
  O(NR^2).
\end{equation}

In our code, instead of the discrete-time Metropolis dynamics we use a rejection free continuous-time version (see \cite{srdinsek_2025} for our implementation). For a configuration $\boldsymbol n$, all possible
single-spin flips are assigned rates
\begin{equation}
  w_i(\boldsymbol n)
  =
  e^{-\Delta_i E^\text{eff}(\boldsymbol n)}.
\end{equation}
 The next spin is sampled from the distribution $w_i/\Gamma$, where $\Gamma(\boldsymbol n)=\sum_i w_i(\boldsymbol n)$,
 and the waiting time is sampled from the exponential law of rate $\Gamma$.
This removes rejections, but after one accepted
flip $j$, the algorithm must update the list of all affected flip rates. Fortunately, only the spins $i$ that interact with $j$ must be updated. There are $O(R^2)$ such atoms with $J^\text{eff}_{ij}\neq 0$.
We get a new energy difference given by
\begin{equation}
\text{new} \ \Delta_i E^\text{eff}(\boldsymbol n) = \begin{cases}
\Delta_i E^\text{eff}(\boldsymbol n) + 2\bar{n}_i \bar{n}_j J_{ij} \text{ if } j\neq i \\
-\Delta_i E^\text{eff}(\boldsymbol n) \text{ if } j= i
\end{cases}
\end{equation}
where the second line comes from the fact that, after flipping $j$, flipping $j$ again just returns to the previous configuration.
 Hence, the $\Delta_i E^\text{eff}$ are cached and with range truncation $R$ its update cost $O(R^2)$ operations.
In order to recover the discrete time Monte Carlo scaling one must cache a data structure that samples and updates the rates $w_i$ without
  rebuilding the full cumulative distribution which, if naively done, costs an $O(N)$ step.
This is achieved by storing the rates in a Fenwick tree \cite{fenwick1994}. After a flip at site $i$, the code
 updates the $O(R^2)$ affected energy differences, and then
updates the corresponding entries of the Fenwick tree. Sampling one spin from
the tree and changing one rate costs $O(\log N)$. In the end, flipping $N$ spins cost $O(NR^2 \log N)$,
 close to the scaling achieved by Hasting-Metropolis algorithm \cref{eq:discrete_scaling}.

\begin{figure}[htbp]
  \centering
  \includegraphics[width=\linewidth]{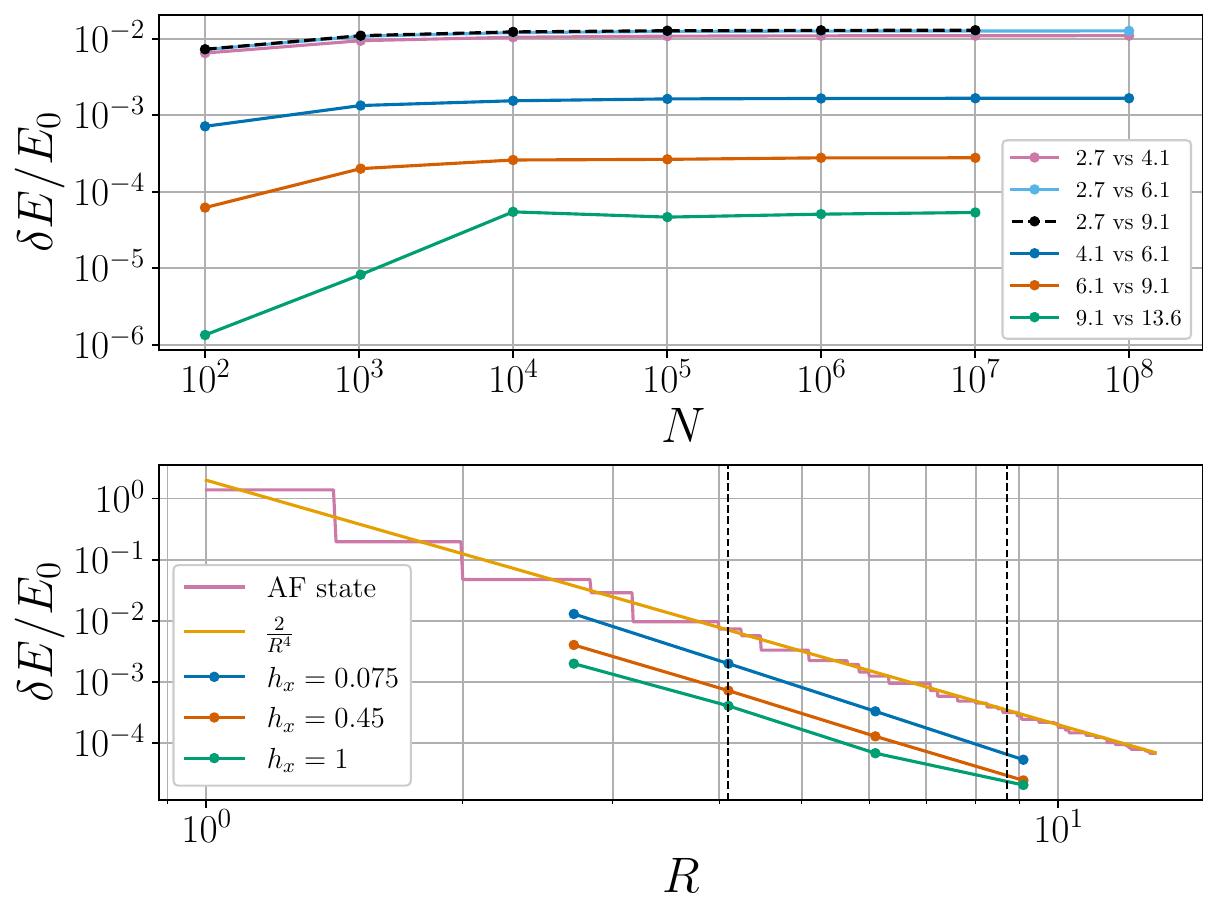}
  \caption{\textbf{Truncation error.} Top: Relative truncation error $|E(R)-E(R')|/E(R)$ that contrasts range $R$ with range $R'>R$ as a function of system size $N$ for different $R$ and $R'$ (see legend), $h_\text{z}=1$ and $h_\text{x}=0.075$.
  Bottom: Relative truncation error versus range $R$.
  Different curves correspond to the $N\rightarrow\infty$, $h_\text{x}=0$, $h_\text{z}=1$ pure antiferromagnetic state for the square lattice (pink line), the $\frac{C}{R^4}$ prediction (orange line) and
 three values of $h_\text{x}$ at $N=10^7$ (blue $h_x=0.075$, red $h_x=0.45$ and green $h_x=1$). The two vertical lines correspond to the range $R=4.1$ and $R=8.7$ used in our large scale simulations.
  All points are calculated with a large number of walkers ensuring a statistical noise level below $10^{-5}$.}
  \label{fig:range_err}
\end{figure}

The last point we need to check is that the truncation induced error is lower than the level of error claimed in \cref{fig:scaling_error}.
The top panel of \cref{fig:range_err} shows the difference in energy between different values of truncations. We find that the rapid decay of the interaction allows us to estimate the truncation error by computing the difference of energy between range $R$ and $R'\approx 1.5R$ (collapse of the top three curves with $R=2.7$ and $R'=4.1$, $R'=6.1$ and $R'=9.1$).
This is shown here at $h_\text{x}=0.075$ and $R=2.7$ for VMC as the three $R=2.7$ curves are superposed.
 Similar behaviors are observed for other values of $R$, $h_\text{x}$ and for GFMC (not shown).
The lower panel of \cref{fig:range_err} shows the classical error for the square lattice anti-ferromagnetic ground state at $h_\text{x}=0$.
 This is the error of
\begin{align}
E(R) &= \sum_{i,|\vec{r}_i|<R} \frac{n_i}{|\vec{r}_i|^6}-h_\text{z} \\
&\approx \int_0^R 2\pi rdr/r^6 - h_\text{z}\\
&\approx E(R=\infty) - \frac{C}{R^4}
\end{align}
 with $C/E(R=\infty) \approx 2$ for $h_\text{z}=1$.
 The three other lines for increasing values of  $h_\text{x}$
 show that the error level stay consistent upon switching on the transverse magnetic field (the error actually appears to decrease slightly as the diagonal energy becomes less and less relevant).
 Interestingly the error at $h_\text{x}>0$ also decays as $1/R^4$.
Conversely, the range truncation error in \cref{fig:range_err} can be seen as a limitation of Rydberg atom based annealers. If one wants to study a given QUBO problem, the $1/r^6$ tails may actually be a problem that limits the accuracy to somewhere in the 1\% level. It is not clear how to alleviate this problem.

%% file: sections/appendix_first_order.tex
\section{Evidence for first- versus second-order transition}\label{app:firstorder}
\begin{figure}[htbp]
  \centering
  \includegraphics[width=\linewidth]{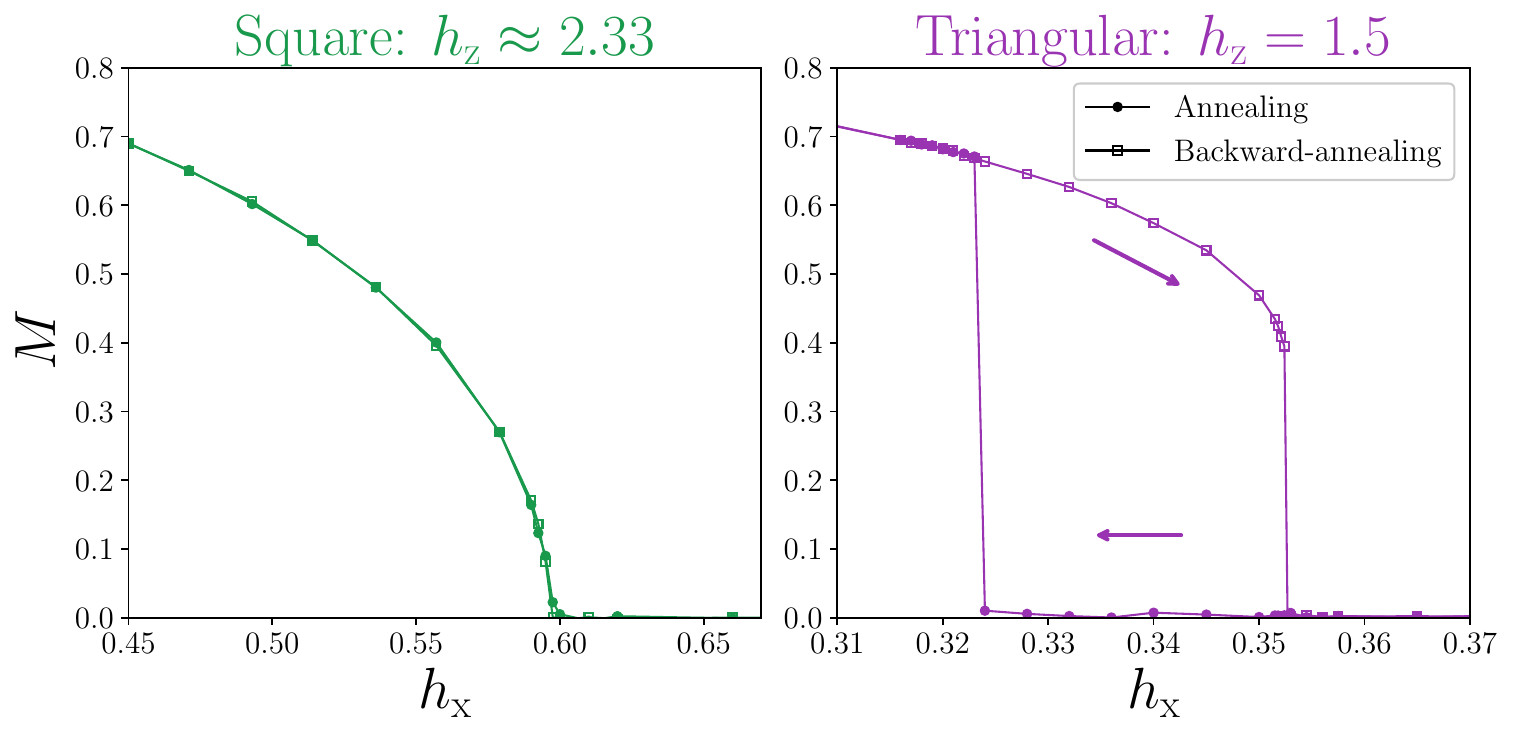}
  \caption{\textbf{First- versus second-order phase transition.} Staggered magnetization $M$ as the transverse field $h_\text{x}$ decreases/increases (annealing/backward-annealing).
  Left panel: square lattice at $h_\text{z}\approx 2.33$ and $N=100$ atoms. Right panel: triangular lattice at $h_\text{z}=1.5$ and $N=147$ atoms.}
  \label{fig:hysteresis}
\end{figure}

\cref{fig:hysteresis} shows a calculation with the staggered thermal ansatz where we have performed a standard annealing (slowly decreasing $h_\text{x}$, optimizing the effective temperatures at every step) as well as a
backward-annealing (slowly increasing $h_\text{x}$, also optimizing the effective temperatures at every step). For the square lattice, the staggered
magnetization shows no hysteresis and a $M(h_\text{x})$ curve consistent with a second order transition. However, for the triangular lattice, we observe jumps of $M$ as well as a hysteresis loop between the forward and backward paths. This is consistent with a first order transition, as predicted previously \cite{janke1997,guo2023}. It is rather appealing that the energy manifold in the parameters $\beta$ has local minima that reflect
the physical structure of the system. Again, we attribute this feature to the fact that the ansatz is physically inspired. We would not expect such a behaviour from, say, a neural quantum state.

%% file: bibliography/references.bib
@article{albash2018,
  title = {Demonstration of a Scaling Advantage for a Quantum Annealer over Simulated Annealing},
  author = {Albash, Tameem and Lidar, Daniel A.},
  journal = {Phys. Rev. X},
  volume = {8},
  issue = {3},
  pages = {031016},
  numpages = {26},
  year = {2018},
  month = {Jul},
  publisher = {American Physical Society},
  doi = {10.1103/PhysRevX.8.031016},
  url = {https://link.aps.org/doi/10.1103/PhysRevX.8.031016}
}

@article{altshuler2010,
author = {Boris Altshuler  and Hari Krovi  and Jérémie Roland },
title = {Anderson localization makes adiabatic quantum optimization fail},
journal = {Proceedings of the National Academy of Sciences},
volume = {107},
number = {28},
pages = {12446-12450},
year = {2010},
doi = {10.1073/pnas.1002116107},
URL = {https://www.pnas.org/doi/abs/10.1073/pnas.1002116107},
eprint = {https://www.pnas.org/doi/pdf/10.1073/pnas.1002116107}}

@article{amin2009,
  title = {First-order quantum phase transition in adiabatic quantum computation},
  author = {Amin, M. H. S. and Choi, V.},
  journal = {Phys. Rev. A},
  volume = {80},
  issue = {6},
  pages = {062326},
  numpages = {5},
  year = {2009},
  month = {Dec},
  publisher = {American Physical Society},
  doi = {10.1103/PhysRevA.80.062326},
  url = {https://link.aps.org/doi/10.1103/PhysRevA.80.062326}
}

@article{ardonne2004,
title = {Topological order and conformal quantum critical points},
journal = {Annals of Physics},
volume = {310},
number = {2},
pages = {493-551},
year = {2004},
issn = {0003-4916},
doi = {https://doi.org/10.1016/j.aop.2004.01.004},
url = {https://www.sciencedirect.com/science/article/pii/S0003491604000247},
author = {Eddy Ardonne and Paul Fendley and Eduardo Fradkin}
}

@article{bapst2013,
	author = {Bapst, V. and Foini, L. and Krzakala, F. and Semerjian, G. and Zamponi, F.},
	title = {{The quantum adiabatic algorithm applied to random optimization problems: The quantum spin glass perspective}},
	journal = {Phys. Rep.},
	volume = {523},
	number = {3},
	pages = {127--205},
	year = {2013},
	month = feb,
	issn = {0370-1573},
	publisher = {North-Holland},
	doi = {10.1016/j.physrep.2012.10.002}
}

@book{becca2017,
  title={Quantum Monte Carlo approaches for correlated systems},
  author={Becca, Federico and Sorella, Sandro},
  year={2017},
  publisher={Cambridge University Press}
}

@article{Benedetti2017,
  title = {Quantum-Assisted Learning of Hardware-Embedded Probabilistic Graphical Models},
  author = {Benedetti, Marcello and Realpe-G\'omez, John and Biswas, Rupak and Perdomo-Ortiz, Alejandro},
  journal = {Phys. Rev. X},
  volume = {7},
  issue = {4},
  pages = {041052},
  numpages = {17},
  year = {2017},
  month = {Nov},
  publisher = {American Physical Society},
  doi = {10.1103/PhysRevX.7.041052},
  url = {https://link.aps.org/doi/10.1103/PhysRevX.7.041052}
}

@article{boixo2014,
	title = {Evidence for quantum annealing with more than one hundred qubits},
	volume = {10},
	issn = {1745-2481},
	url = {https://doi.org/10.1038/nphys2900},
	doi = {10.1038/nphys2900},
	number = {3},
	journal = {Nature Physics},
	author = {Boixo, Sergio and Rønnow, Troels F. and Isakov, Sergei V. and Wang, Zhihui and Wecker, David and Lidar, Daniel A. and Martinis, John M. and Troyer, Matthias},
	month = mar,
	year = {2014},
	pages = {218--224},
}

@article{bombieri2025,
  title = {Deconfined Quantum Criticality on a Triangular Rydberg Array},
  author = {Bombieri, Lisa and Zache, Torsten V. and Calliari, Gabriele and Lukin, Mikhail D. and Pichler, Hannes and Gonz\'alez-Cuadra, Daniel},
  journal = {Phys. Rev. Lett.},
  volume = {135},
  issue = {23},
  pages = {233602},
  numpages = {7},
  year = {2025},
  month = {Dec},
  publisher = {American Physical Society},
  doi = {10.1103/4rg4-zbnn},
  url = {https://link.aps.org/doi/10.1103/4rg4-zbnn}
}

@article{brodoloni2026,
	author = {Brodoloni, L. and Astrakharchik, G. E. and Giorgini, S. and Pilati, S.},
	title = {{Energy gap of quantum spin glasses: a projection quantum Monte Carlo study}},
	journal = {arXiv},
	year = {2026},
	month = feb,
	eprint = {2602.20108},
	doi = {https://doi.org/10.48550/arXiv.2602.20108}
}

@article{carugno2022,
	author = {Carugno, Costantino and Ferrari Dacrema, Maurizio and Cremonesi, Paolo},
	title = {{Evaluating the job shop scheduling problem on a D-wave quantum annealer}},
	journal = {Sci. Rep.},
	volume = {12},
	number = {6539},
	pages = {6539},
	year = {2022},
	month = apr,
	issn = {2045-2322},
	publisher = {Nature Publishing Group},
	doi = {10.1038/s41598-022-10169-0}
}

@article{castelnovo2005,
	author = {Castelnovo, Claudio and Chamon, Claudio and Mudry, Christopher and Pujol, Pierre},
	title = {{From quantum mechanics to classical statistical physics: Generalized Rokhsar{\textendash}Kivelson Hamiltonians and the {\textquotedblleft}Stochastic Matrix Form{\textquotedblright} decomposition}},
	journal = {Ann. Phys.},
	volume = {318},
	number = {2},
	pages = {316--344},
	year = {2005},
	month = aug,
	issn = {0003-4916},
	publisher = {Academic Press},
	doi = {10.1016/j.aop.2005.01.006}
}

@article{cazals2025,
	author = {Cazals, Pierre and Fran{\ifmmode\mbox{\c{c}}\else\c{c}\fi}ois, Aymeric and Henriet, Lo{\ifmmode\ddot{\imath}\else\"{\i}\fi}c and Leclerc, Lucas and Marin, Malory and Naghmouchi, Yassine and Coelho, Wesley da Silva and Sikora, Florian and Vitale, Vittorio and Watrigant, R{\ifmmode\acute{e}\else\'{e}\fi}mi and Garzillo, Monique Witt and Dalyac, Constantin},
	title = {{Identifying hard native instances for the maximum-independent-set problem on neutral-atom quantum processors}},
	journal = {Phys. Rev. Appl.},
	volume = {25},
	number = {3},
	pages = {034085},
	year = {2026},
	month = mar,
	publisher = {American Physical Society},
	doi = {10.1103/8gjv-ll1f}
}

@article{ceperley1995,
	author = {Ceperley, D. M.},
	title = {{Path integrals in the theory of condensed helium}},
	journal = {Rev. Mod. Phys.},
	volume = {67},
	number = {2},
	pages = {279--355},
	year = {1995},
	month = apr,
	publisher = {American Physical Society},
	doi = {10.1103/RevModPhys.67.279}
}

@article{denchev2016,
	author = {Denchev, Vasil S. and Boixo, Sergio and Isakov, Sergei V. and Ding, Nan and Babbush, Ryan and Smelyanskiy, Vadim and Martinis, John and Neven, Hartmut},
	title = {{What is the Computational Value of Finite-Range Tunneling?}},
	journal = {Phys. Rev. X},
	volume = {6},
	number = {3},
	pages = {031015},
	year = {2016},
	month = aug,
	publisher = {American Physical Society},
	doi = {10.1103/PhysRevX.6.031015}
}

@article{ebadi2021,
	author = {Ebadi, Sepehr and Wang, Tout T. and Levine, Harry and Keesling, Alexander and Semeghini, Giulia and Omran, Ahmed and Bluvstein, Dolev and Samajdar, Rhine and Pichler, Hannes and Ho, Wen Wei and Choi, Soonwon and Sachdev, Subir and Greiner, Markus and Vuleti{\ifmmode\acute{c}\else\'{c}\fi}, Vladan and Lukin, Mikhail D.},
	title = {{Quantum phases of matter on a 256-atom programmable quantum simulator}},
	journal = {Nature},
	volume = {595},
	pages = {227--232},
	year = {2021},
	month = jul,
	issn = {1476-4687},
	publisher = {Nature Publishing Group},
	doi = {10.1038/s41586-021-03582-4}
}

@article{ebadi2022,
	author = {Ebadi, S. and Keesling, A. and Cain, M. and Wang, T. T. and Levine, H. and Bluvstein, D. and Semeghini, G. and Omran, A. and Liu, J.-G. and Samajdar, R. and Luo, X.-Z. and Nash, B. and Gao, X. and Barak, B. and Farhi, E. and Sachdev, S. and Gemelke, N. and Zhou, L. and Choi, S. and Pichler, H. and Wang, S.-T. and Greiner, M. and Vuleti{\ifmmode\acute{c}\else\'{c}\fi}, V. and Lukin, M. D.},
	title = {{Quantum optimization of maximum independent set using Rydberg atom arrays}},
	journal = {Science},
	volume = {376},
	number = {6598},
	pages = {1209--1215},
	year = {2022},
	month = may,
	issn = {0036-8075},
	publisher = {American Association for the Advancement of Science},
	doi = {10.1126/science.abo6587}
}

@article{farhi2011,
	author = {Farhi, Edward and Goldstone, Jeffrey and Gosset, David and Gutmann, Sam and Meyer, Harvey B. and Shor, Peter},
	title = {{Quantum Adiabatic Algorithms, Small Gaps, and Different Paths}},
	journal = {arXiv},
	year = {2009},
	month = sep,
	eprint = {0909.4766},
	doi = {10.48550/arXiv.0909.4766}
}

@article{fenwick1994,
	author = {Fenwick, Peter M.},
	title = {{A new data structure for cumulative frequency tables}},
	journal = {Softw.: Pract. Exper.},
	volume = {24},
	number = {3},
	pages = {327--336},
	year = {1994},
	month = mar,
	issn = {0038-0644},
	publisher = {John Wiley {\&} Sons, Ltd},
	doi = {10.1002/spe.4380240306},
	url = {https://onlinelibrary.wiley.com/doi/abs/10.1002/spe.4380240306}
}

@article{fey2019,
	author = {Fey, Sebastian and Kapfer, Sebastian C. and Schmidt, Kai Phillip},
	title = {{Quantum Criticality of Two-Dimensional Quantum Magnets with Long-Range Interactions}},
	journal = {Phys. Rev. Lett.},
	volume = {122},
	number = {1},
	pages = {017203},
	year = {2019},
	month = jan,
	publisher = {American Physical Society},
	doi = {10.1103/PhysRevLett.122.017203}
}

@article{guo2023,
	author = {Guo, Sibo and Huang, Juntao and Hu, Jiangping and Li, Zi-Xiang},
	title = {{Order by disorder and an emergent Kosterlitz-Thouless phase in a triangular Rydberg array}},
	journal = {Phys. Rev. A},
	volume = {108},
	number = {5},
	pages = {053314},
	year = {2023},
	month = nov,
	publisher = {American Physical Society},
	doi = {10.1103/PhysRevA.108.053314}
}

@article{hajek1988,
  title={Cooling schedules for optimal annealing},
  author={Hajek, B.},
  journal={Mathematics of Operations Research},
  volume={13},
  number={2},
  pages={311--329},
  year={1988},
  url={https://web.mit.edu/6.435/www/Hajek88.pdf}
}

@article{heim2015,
	author = {Heim, Bettina and R{\o}nnow, Troels F. and Isakov, Sergei V. and Troyer, Matthias},
	title = {{Quantum versus classical annealing of Ising spin glasses}},
	journal = {Science},
	volume = {348},
	number = {6231},
	pages = {215--217},
	year = {2015},
	month = mar,
	issn = {0036-8075},
	publisher = {American Association for the Advancement of Science},
	doi = {10.1126/science.aaa4170}
}

@article{henley2004,
	author = {Henley, C. L.},
	title = {{From classical to quantum dynamics at Rokhsar{\textendash}Kivelson points}},
	journal = {J. Phys.: Condens. Matter},
	volume = {16},
	number = {11},
	pages = {S891},
	year = {2004},
	month = mar,
	issn = {0953-8984},
	publisher = {IOP Publishing},
	doi = {10.1088/0953-8984/16/11/045}
}

@article{inagaki2016,
	author = {Inagaki, Takahiro and Haribara, Yoshitaka and Igarashi, Koji and Sonobe, Tomohiro and Tamate, Shuhei and Honjo, Toshimori and Marandi, Alireza and McMahon, Peter L. and Umeki, Takeshi and Enbutsu, Koji and Tadanaga, Osamu and Takenouchi, Hirokazu and Aihara, Kazuyuki and Kawarabayashi, Ken-ichi and Inoue, Kyo and Utsunomiya, Shoko and Takesue, Hiroki},
	title = {{A coherent Ising machine for 2000-node optimization problems}},
	journal = {Science},
	volume = {354},
	number = {6312},
	pages = {603--606},
	year = {2016},
	month = oct,
	issn = {0036-8075},
	publisher = {American Association for the Advancement of Science},
	doi = {10.1126/science.aah4243}
}

@article{isakov2003,
	author = {Isakov, S. V. and Moessner, R.},
	title = {{Interplay of quantum and thermal fluctuations in a frustrated magnet}},
	journal = {Phys. Rev. B},
	volume = {68},
	number = {10},
	pages = {104409},
	year = {2003},
	month = sep,
	publisher = {American Physical Society},
	doi = {10.1103/PhysRevB.68.104409}
}

@article{janke1997,
title = {Three-dimensional 3-state Potts model revisited with new techniques},
journal = {Nuclear Physics B},
volume = {489},
number = {3},
pages = {679-696},
year = {1997},
issn = {0550-3213},
doi = {https://doi.org/10.1016/S0550-3213(96)00710-9},
url = {https://www.sciencedirect.com/science/article/pii/S0550321396007109},
author = {Wolfhard Janke and Ramon Villanova},
}

@article{johnson2011,
	title = {Quantum annealing with manufactured spins},
	volume = {473},
	issn = {1476-4687},
	url = {https://doi.org/10.1038/nature10012},
	doi = {10.1038/nature10012},
	number = {7346},
	journal = {Nature},
	author = {Johnson, M. W. and Amin, M. H. S. and Gildert, S. and Lanting, T. and Hamze, F. and Dickson, N. and Harris, R. and Berkley, A. J. and Johansson, J. and Bunyk, P. and Chapple, E. M. and Enderud, C. and Hilton, J. P. and Karimi, K. and Ladizinsky, E. and Ladizinsky, N. and Oh, T. and Perminov, I. and Rich, C. and Thom, M. C. and Tolkacheva, E. and Truncik, C. J. S. and Uchaikin, S. and Wang, J. and Wilson, B. and Rose, G.},
	month = may,
	year = {2011},
	pages = {194--198},
}

@article{jorg2008,
	author = {J{\ifmmode\ddot{o}\else\"{o}\fi}rg, Thomas and Krzakala, Florent and Kurchan, Jorge and Maggs, A. C.},
	title = {{Simple Glass Models and Their Quantum Annealing}},
	journal = {Phys. Rev. Lett.},
	volume = {101},
	number = {14},
	pages = {147204},
	year = {2008},
	month = oct,
	publisher = {American Physical Society},
	doi = {10.1103/PhysRevLett.101.147204}
}

@article{jorg2010,
  title = {First-Order Transitions and the Performance of Quantum Algorithms in Random Optimization Problems},
  author = {J\"org, Thomas and Krzakala, Florent and Semerjian, Guilhem and Zamponi, Francesco},
  journal = {Phys. Rev. Lett.},
  volume = {104},
  issue = {20},
  pages = {207206},
  numpages = {4},
  year = {2010},
  month = {May},
  publisher = {American Physical Society},
  doi = {10.1103/PhysRevLett.104.207206},
  url = {https://link.aps.org/doi/10.1103/PhysRevLett.104.207206}
}

@article{kadowaki1998,
	author = {Kadowaki, Tadashi and Nishimori, Hidetoshi},
	title = {{Quantum annealing in the transverse Ising model}},
	journal = {Phys. Rev. E},
	volume = {58},
	number = {5},
	pages = {5355--5363},
	year = {1998},
	month = nov,
	publisher = {American Physical Society},
	doi = {10.1103/PhysRevE.58.5355}
}

@article{kalinowski_2022,
	title = {Bulk and {Boundary} {Quantum} {Phase} {Transitions} in a {Square} {Rydberg} {Atom} {Array}},
	volume = {105},
	issn = {2469-9950, 2469-9969},
	url = {http://arxiv.org/abs/2112.10790},
	doi = {10.1103/PhysRevB.105.174417},
	number = {17},
	urldate = {2025-08-14},
	journal = {Physical Review B},
	author = {Kalinowski, Marcin and Samajdar, Rhine and Melko, Roger G. and Lukin, Mikhail D. and Sachdev, Subir and Choi, Soonwon},
	month = may,
	year = {2022},
	note = {arXiv:2112.10790 [quant-ph]},
	pages = {174417},
}

@article{kashima2001,
	author = {Kashima, Tsuyoshi and Imada, Masatoshi},
	title = {{Path-Integral Renormalization Group Method for Numerical Study on Ground States of Strongly Correlated Electronic Systems}},
	journal = {J. Phys. Soc. Jpn.},
	volume = {70},
	number = {8},
	pages = {2287--2299},
	year = {2001},
	month = aug,
	issn = {0031-9015},
	publisher = {The Physical Society of Japan},
	doi = {10.1143/JPSJ.70.2287}
}

@article{King2018,
	author = {King, Andrew D. and Carrasquilla, Juan and Raymond, Jack and Ozfidan, Isil and Andriyash, Evgeny and Berkley, Andrew and Reis, Mauricio and Lanting, Trevor and Harris, Richard and Altomare, Fabio and Boothby, Kelly and Bunyk, Paul I. and Enderud, Colin and Fr{\ifmmode\acute{e}\else\'{e}\fi}chette, Alexandre and Hoskinson, Emile and Ladizinsky, Nicolas and Oh, Travis and Poulin-Lamarre, Gabriel and Rich, Christopher and Sato, Yuki and Smirnov, Anatoly {\relax Yu}. and Swenson, Loren J. and Volkmann, Mark H. and Whittaker, Jed and Yao, Jason and Ladizinsky, Eric and Johnson, Mark W. and Hilton, Jeremy and Amin, Mohammad H.},
	title = {{Observation of topological phenomena in a programmable lattice of 1,800 qubits}},
	journal = {Nature},
	volume = {560},
	pages = {456--460},
	year = {2018},
	month = aug,
	issn = {1476-4687},
	publisher = {Nature Publishing Group},
	doi = {10.1038/s41586-018-0410-x}
}

@article{King2021,
	author = {King, Andrew D. and Raymond, Jack and Lanting, Trevor and Isakov, Sergei V. and Mohseni, Masoud and Poulin-Lamarre, Gabriel and Ejtemaee, Sara and Bernoudy, William and Ozfidan, Isil and Smirnov, Anatoly {\relax Yu}. and Reis, Mauricio and Altomare, Fabio and Babcock, Michael and Baron, Catia and Berkley, Andrew J. and Boothby, Kelly and Bunyk, Paul I. and Christiani, Holly and Enderud, Colin and Evert, Bram and Harris, Richard and Hoskinson, Emile and Huang, Shuiyuan and Jooya, Kais and Khodabandelou, Ali and Ladizinsky, Nicolas and Li, Ryan and Lott, P. Aaron and MacDonald, Allison J. R. and Marsden, Danica and Marsden, Gaelen and Medina, Teresa and Molavi, Reza and Neufeld, Richard and Norouzpour, Mana and Oh, Travis and Pavlov, Igor and Perminov, Ilya and Prescott, Thomas and Rich, Chris and Sato, Yuki and Sheldan, Benjamin and Sterling, George and Swenson, Loren J. and Tsai, Nicholas and Volkmann, Mark H. and Whittaker, Jed D. and Wilkinson, Warren and Yao, Jason and Neven, Hartmut and Hilton, Jeremy P. and Ladizinsky, Eric and Johnson, Mark W. and Amin, Mohammad H.},
	title = {{Scaling advantage over path-integral Monte Carlo in quantum simulation of geometrically frustrated magnets}},
	journal = {Nat. Commun.},
	volume = {12},
	number = {1113},
	pages = {1113},
	year = {2021},
	month = feb,
	issn = {2041-1723},
	publisher = {Nature Publishing Group},
	doi = {10.1038/s41467-021-20901-5}
}

@article{King2022,
	author = {King, Andrew D. and Suzuki, Sei and Raymond, Jack and Zucca, Alex and Lanting, Trevor and Altomare, Fabio and Berkley, Andrew J. and Ejtemaee, Sara and Hoskinson, Emile and Huang, Shuiyuan and Ladizinsky, Eric and MacDonald, Allison J. R. and Marsden, Gaelen and Oh, Travis and Poulin-Lamarre, Gabriel and Reis, Mauricio and Rich, Chris and Sato, Yuki and Whittaker, Jed D. and Yao, Jason and Harris, Richard and Lidar, Daniel A. and Nishimori, Hidetoshi and Amin, Mohammad H.},
	title = {{Coherent quantum annealing in a programmable 2,000{\hspace{0.167em}}qubit Ising chain}},
	journal = {Nat. Phys.},
	volume = {18},
	pages = {1324--1328},
	year = {2022},
	month = nov,
	issn = {1745-2481},
	publisher = {Nature Publishing Group},
	doi = {10.1038/s41567-022-01741-6}
}

@article{King2023,
	author = {King, Andrew D. and Raymond, Jack and Lanting, Trevor and Harris, Richard and Zucca, Alex and Altomare, Fabio and Berkley, Andrew J. and Boothby, Kelly and Ejtemaee, Sara and Enderud, Colin and Hoskinson, Emile and Huang, Shuiyuan and Ladizinsky, Eric and MacDonald, Allison J. R. and Marsden, Gaelen and Molavi, Reza and Oh, Travis and Poulin-Lamarre, Gabriel and Reis, Mauricio and Rich, Chris and Sato, Yuki and Tsai, Nicholas and Volkmann, Mark and Whittaker, Jed D. and Yao, Jason and Sandvik, Anders W. and Amin, Mohammad H.},
	title = {{Quantum critical dynamics in a 5,000-qubit programmable spin glass}},
	journal = {Nature},
	volume = {617},
	pages = {61--66},
	year = {2023},
	month = may,
	issn = {1476-4687},
	publisher = {Nature Publishing Group},
	doi = {10.1038/s41586-023-05867-2}
}

@article{king2025,
	author = {King, Andrew D. and Nocera, Alberto and Rams, Marek M. and Dziarmaga, Jacek and Wiersema, Roeland and Bernoudy, William and Raymond, Jack and Kaushal, Nitin and Heinsdorf, Niclas and Harris, Richard and Boothby, Kelly and Altomare, Fabio and Asad, Mohsen and Berkley, Andrew J. and Boschnak, Martin and Chern, Kevin and Christiani, Holly and Cibere, Samantha and Connor, Jake and Dehn, Martin H. and Deshpande, Rahul and Ejtemaee, Sara and Farre, Pau and Hamer, Kelsey and Hoskinson, Emile and Huang, Shuiyuan and Johnson, Mark W. and Kortas, Samuel and Ladizinsky, Eric and Lanting, Trevor and Lai, Tony and Li, Ryan and MacDonald, Allison J. R. and Marsden, Gaelen and McGeoch, Catherine C. and Molavi, Reza and Oh, Travis and Neufeld, Richard and Norouzpour, Mana and Pasvolsky, Joel and Poitras, Patrick and Poulin-Lamarre, Gabriel and Prescott, Thomas and Reis, Mauricio and Rich, Chris and Samani, Mohammad and Sheldan, Benjamin and Smirnov, Anatoly and Sterpka, Edward and Clavera, Berta Trullas and Tsai, Nicholas and Volkmann, Mark and Whiticar, Alexander M. and Whittaker, Jed D. and Wilkinson, Warren and Yao, Jason and Yi, T. J. and Sandvik, Anders W. and Alvarez, Gonzalo and Melko, Roger G. and Carrasquilla, Juan and Franz, Marcel and Amin, Mohammad H.},
	title = {{Beyond-classical computation in quantum simulation}},
	journal = {Science},
	volume = {388},
	number = {6743},
	pages = {199--204},
	year = {2025},
	month = mar,
	issn = {0036-8075},
	publisher = {American Association for the Advancement of Science},
	doi = {10.1126/science.ado6285}
}

@article{kirkpatrick1983,
  title={Optimization by simulated annealing},
  author={Kirkpatrick, S. and Gelatt, C.D. Jr. and Vecchi, M.P.},
  journal={Science},
  volume={220},
  number={4598},
  pages={671--680},
  year={1983},
  doi={10.1126/science.220.4598.671}
}

@article{kochenberger2014,
	author = {Kochenberger, Gary and Hao, Jin-Kao and Glover, Fred and Lewis, Mark and L{\ifmmode\ddot{u}\else\"{u}\fi}, Zhipeng and Wang, Haibo and Wang, Yang},
	title = {{The unconstrained binary quadratic programming problem: a survey}},
	journal = {J. Comb. Optim.},
	volume = {28},
	number = {1},
	pages = {58--81},
	year = {2014},
	month = jul,
	issn = {1573-2886},
	publisher = {Springer US},
	doi = {10.1007/s10878-014-9734-0}
}

@article{konz2021,
	author = {K{\ifmmode\ddot{o}\else\"{o}\fi}nz, Mario S. and Lechner, Wolfgang and Katzgraber, Helmut G. and Troyer, Matthias},
	title = {{Embedding Overhead Scaling of Optimization Problems in Quantum Annealing}},
	journal = {PRX Quantum},
	volume = {2},
	number = {4},
	pages = {040322},
	year = {2021},
	month = nov,
	publisher = {American Physical Society},
	doi = {10.1103/PRXQuantum.2.040322}
}

@article{koziol2019,
	author = {Koziol, Jan and Fey, Sebastian and Kapfer, Sebastian C. and Schmidt, Kai Phillip},
	title = {{Quantum criticality of the transverse-field Ising model with long-range interactions on triangular-lattice cylinders}},
	journal = {Phys. Rev. B},
	volume = {100},
	number = {14},
	pages = {144411},
	year = {2019},
	month = oct,
	publisher = {American Physical Society},
	doi = {10.1103/PhysRevB.100.144411}
}

@article{lanting2014,
	author = {Lanting, T. and Przybysz, A. J. and Smirnov, A. {\relax Yu}. and Spedalieri, F. M. and Amin, M. H. and Berkley, A. J. and Harris, R. and Altomare, F. and Boixo, S. and Bunyk, P. and Dickson, N. and Enderud, C. and Hilton, J. P. and Hoskinson, E. and Johnson, M. W. and Ladizinsky, E. and Ladizinsky, N. and Neufeld, R. and Oh, T. and Perminov, I. and Rich, C. and Thom, M. C. and Tolkacheva, E. and Uchaikin, S. and Wilson, A. B. and Rose, G.},
	title = {{Entanglement in a Quantum Annealing Processor}},
	journal = {Phys. Rev. X},
	volume = {4},
	number = {2},
	pages = {021041},
	year = {2014},
	month = may,
	publisher = {American Physical Society},
	doi = {10.1103/PhysRevX.4.021041}
}

@article{laumann2012,
	author = {Laumann, C. R. and Moessner, R. and Scardicchio, A. and Sondhi, S. L.},
	title = {{Quantum Adiabatic Algorithm and Scaling of Gaps at First-Order Quantum Phase Transitions}},
	journal = {Phys. Rev. Lett.},
	volume = {109},
	number = {3},
	pages = {030502},
	year = {2012},
	month = jul,
	publisher = {American Physical Society},
	doi = {10.1103/PhysRevLett.109.030502}
}

@article{leclerc2022financial,
	author = {Leclerc, Lucas and Ortiz-Guti{\ifmmode\acute{e}\else\'{e}\fi}rrez, Luis and Grijalva, Sebasti{\ifmmode\acute{a}\else\'{a}\fi}n and Albrecht, Boris and Cline, Julia R. K. and Elfving, Vincent E. and Signoles, Adrien and Henriet, Lo{\ifmmode\ddot{\imath}\else\"{\i}\fi}c and Del Bimbo, Gianni and Sheikh, Usman Ayub and Shah, Maitree and Andrea, Luc and Ishtiaq, Faysal and Duarte, Andoni and Mugel, Sam and C{\ifmmode\acute{a}\else\'{a}\fi}ceres, Irene and Kurek, Michel and Or{\ifmmode\acute{u}\else\'{u}\fi}s, Roman and Seddik, Achraf and Hammami, Oumaima and Isselnane, Hacene and M'tamon, Didier},
	title = {{Financial risk management on a neutral atom quantum processor}},
	journal = {Phys. Rev. Res.},
	volume = {5},
	number = {4},
	pages = {043117},
	year = {2023},
	month = nov,
	publisher = {American Physical Society},
	doi = {10.1103/PhysRevResearch.5.043117}
}

@article{louvet2026,
  title = {Feasibility of performing quantum chemistry calculations on quantum computers},
  author = {Louvet, Thibaud and Ayral, Thomas and Waintal, Xavier},
  journal = {Phys. Rev. B},
  volume = {113},
  issue = {12},
  pages = {125112},
  numpages = {15},
  year = {2026},
  month = {Mar},
  publisher = {American Physical Society},
  doi = {10.1103/hpt6-9tnk},
  url = {https://link.aps.org/doi/10.1103/hpt6-9tnk}
}

@article{luchnikov2024,
	author = {Luchnikov, Ilia A. and Tiunov, Egor S. and Haug, Tobias and Aolita, Leandro},
	title = {{Large-scale quantum annealing simulation with tensor networks and belief propagation}},
	journal = {arXiv},
	year = {2024},
	month = sep,
	eprint = {2409.12240},
	doi = {10.48550/arXiv.2409.12240}
}

@article{martonak2004,
  title = {Quantum annealing of the traveling-salesman problem},
  author = {Marto\ifmmode \check{n}\else \v{n}\fi{}\'ak, Roman and Santoro, Giuseppe E. and Tosatti, Erio},
  journal = {Phys. Rev. E},
  volume = {70},
  issue = {5},
  pages = {057701},
  numpages = {4},
  year = {2004},
  month = {Nov},
  publisher = {American Physical Society},
  doi = {10.1103/PhysRevE.70.057701},
  url = {https://link.aps.org/doi/10.1103/PhysRevE.70.057701}
}

@article{mauron2025,
	author = {Mauron, Linda and Carleo, Giuseppe},
	title = {{Challenging the Quantum Advantage Frontier with Large-Scale Classical Simulations of Annealing Dynamics}},
	journal = {arXiv},
	year = {2025},
	month = mar,
	eprint = {2503.08247},
	doi = {10.48550/arXiv.2503.08247}
}

@article{miller1993,
	author = {Miller, Jonathan and Huse, David A.},
	title = {{Zero-temperature critical behavior of the infinite-range quantum Ising spin glass}},
	journal = {Phys. Rev. Lett.},
	volume = {70},
	number = {20},
	pages = {3147--3150},
	year = {1993},
	month = may,
	publisher = {American Physical Society},
	doi = {10.1103/PhysRevLett.70.3147}
}

@article{moessner1999,
	author = {Moessner, R. and Sondhi, S. L. and Chandra, P.},
	title = {{Two-Dimensional Periodic Frustrated Ising Models in a Transverse Field}},
	journal = {Phys. Rev. Lett.},
	volume = {84},
	number = {19},
	pages = {4457--4460},
	year = {2000},
	month = may,
	publisher = {American Physical Society},
	doi = {10.1103/PhysRevLett.84.4457}
}

@article{moessner2001,
	author = {Moessner, R. and Sondhi, S. L.},
	title = {{Ising models of quantum frustration}},
	journal = {Phys. Rev. B},
	volume = {63},
	number = {22},
	pages = {224401},
	year = {2001},
	month = may,
	publisher = {American Physical Society},
	doi = {10.1103/PhysRevB.63.224401}
}

@article{mohseni2022,
	title = {Ising machines as hardware solvers of combinatorial optimization problems},
	volume = {4},
	issn = {2522-5820},
	url = {https://doi.org/10.1038/s42254-022-00440-8},
	doi = {10.1038/s42254-022-00440-8},
	number = {6},
	journal = {Nature Reviews Physics},
	author = {Mohseni, Naeimeh and McMahon, Peter L. and Byrnes, Tim},
	month = jun,
	year = {2022},
	pages = {363--379},
}

@article{mora2007,
	author = {Mora, Christophe and Waintal, Xavier},
	title = {{Variational Wave Functions and Their Overlap with the Ground State}},
	journal = {Phys. Rev. Lett.},
	volume = {99},
	number = {3},
	pages = {030403},
	year = {2007},
	month = jul,
	publisher = {American Physical Society},
	doi = {10.1103/PhysRevLett.99.030403}
}

@article{morita2008,
	author = {Morita, Satoshi and Nishimori, Hidetoshi},
	title = {{Mathematical foundation of quantum annealing}},
	journal = {J. Math. Phys.},
	volume = {49},
	number = {12},
	pages = {125210},
	year = {2008},
	month = dec,
	issn = {0022-2488},
	publisher = {AIP Publishing},
	doi = {10.1063/1.2995837}
}

@article{miyazaki2013,
	author = {Miyazaki, Ryoji and Nishimori, Hidetoshi},
	title = {{Real-space renormalization-group approach to the random transverse-field Ising model in finite dimensions}},
	journal = {Phys. Rev. E},
	volume = {87},
	number = {3},
	pages = {032154},
	year = {2013},
	month = mar,
	publisher = {American Physical Society},
	doi = {10.1103/PhysRevE.87.032154}
}

@Article{naumann_rizzi_introduction_2024,
	title={{An introduction to infinite projected entangled-pair state methods for variational ground state simulations using automatic differentiation}},
	author={Jan Naumann and Erik Lennart Weerda and Matteo Rizzi and Jens Eisert and Philipp Schmoll},
	journal={SciPost Phys. Lect. Notes},
	pages={86},
	year={2024},
	publisher={SciPost},
	doi={10.21468/SciPostPhysLectNotes.86},
	url={https://scipost.org/10.21468/SciPostPhysLectNotes.86}
}

@article{nishimori1996,
author = {Nishimori ,Hidetoshi and Nonomura ,Yoshihiko},
title = {Quantum Effects in Neural Networks},
journal = {Journal of the Physical Society of Japan},
volume = {65},
number = {12},
pages = {3780-3796},
year = {1996},
doi = {10.1143/JPSJ.65.3780},
URL = {         https://doi.org/10.1143/JPSJ.65.3780},
eprint = {         https://doi.org/10.1143/JPSJ.65.3780}
}

@article{ohzeki2011,
	author = {Ohzeki, Masayuki and Nishimori, Hidetoshi},
	title = {{Quantum annealing: An introduction and new developments}},
	journal = {arXiv},
	year = {2010},
	month = jun,
	eprint = {1006.1696},
	doi = {10.48550/arXiv.1006.1696}
}

@article{orus2019,
  title = {Forecasting financial crashes with quantum computing},
  author = {Or\'us, Rom\'an and Mugel, Samuel and Lizaso, Enrique},
  journal = {Phys. Rev. A},
  volume = {99},
  issue = {6},
  pages = {060301},
  numpages = {6},
  year = {2019},
  month = {Jun},
  publisher = {American Physical Society},
  doi = {10.1103/PhysRevA.99.060301},
  url = {https://link.aps.org/doi/10.1103/PhysRevA.99.060301}
}

@misc{phillipson2025,
      title={Quantum Computing in Logistics and Supply Chain Management an Overview}, 
      author={Frank Phillipson},
      year={2025},
      eprint={2402.17520},
      archivePrefix={arXiv},
      primaryClass={quant-ph},
      url={https://arxiv.org/abs/2402.17520}, 
}

@misc{qian2026,
      title={Large Language Model for Discrete Optimization Problems: Evaluation and Step-by-step Reasoning}, 
      author={Tianhao Qian and Guilin Qi and Z. Y. Wu and Ran Gu and Xuanyi Liu and Canchen Lyu},
      year={2026},
      eprint={2603.07733},
      archivePrefix={arXiv},
      primaryClass={cs.AI},
      url={https://arxiv.org/abs/2603.07733}, 
}

@article{quinton2025,
	title = {Quantum annealing applications, challenges and limitations for optimisation problems compared to classical solvers},
	volume = {15},
	issn = {2045-2322},
	url = {https://doi.org/10.1038/s41598-025-96220-2},
	doi = {10.1038/s41598-025-96220-2},
	number = {1},
	journal = {Scientific Reports},
	author = {Quinton, Finley Alexander and Myhr, Per Arne Sevle and Barani, Mostafa and Crespo del Granado, Pedro and Zhang, Hongyu},
	month = apr,
	year = {2025},
	pages = {12733},
}

@article{Raymond2020,
	author = {Raymond, Jack and Yarkoni, Sheir and Andriyash, Evgeny},
	title = {{Global Warming: Temperature Estimation in Annealers}},
	journal = {Front. ICT},
	volume = {3},
	pages = {214379},
	year = {2016},
	month = nov,
	issn = {2297-198X},
	publisher = {Frontiers},
	doi = {10.3389/fict.2016.00023}
}

@inproceedings{reichardt2004, 
	author = {Reichardt, Ben W.}, 
	title = {The quantum adiabatic optimization algorithm and local minima}, 
	year = {2004}, 
	isbn = {1581138520}, 
	publisher = {Association for Computing Machinery}, 
	address = {New York, NY, USA}, 
	url = {https://doi.org/10.1145/1007352.1007428}, 
	doi = {10.1145/1007352.1007428}, 
	booktitle = {Proceedings of the Thirty-Sixth Annual ACM Symposium on Theory of Computing}, 
	pages = {502–510}, 
	numpages = {9}, 
	location = {Chicago, IL, USA}, 
	series = {STOC '04} 
}

@misc{saket2013,
      title={A PTAS for the Classical Ising Spin Glass Problem on the Chimera Graph Structure}, 
      author={Rishi Saket},
      year={2013},
      eprint={1306.6943},
      archivePrefix={arXiv},
      primaryClass={cs.DS},
      url={https://arxiv.org/abs/1306.6943}, 
}

@article{sandvik2003,
  title = {Stochastic series expansion method for quantum Ising models with arbitrary interactions},
  author = {Sandvik, Anders W.},
  journal = {Phys. Rev. E},
  volume = {68},
  issue = {5},
  pages = {056701},
  numpages = {9},
  year = {2003},
  month = {Nov},
  publisher = {American Physical Society},
  doi = {10.1103/PhysRevE.68.056701},
  url = {https://link.aps.org/doi/10.1103/PhysRevE.68.056701}
}

@article{santoro2002,
	author = {Santoro, Giuseppe E. and Marto{\ifmmode\check{n}\else\v{n}\fi}{\ifmmode\acute{a}\else\'{a}\fi}k, Roman and Tosatti, Erio and Car, Roberto},
	title = {{Theory of Quantum Annealing of an Ising Spin Glass}},
	journal = {Science},
	volume = {295},
	number = {5564},
	pages = {2427--2430},
	year = {2002},
	month = mar,
	issn = {0036-8075},
	publisher = {American Association for the Advancement of Science},
	doi = {10.1126/science.1068774}
}

@article{seki2012,
	author = {Seki, Yuya and Nishimori, Hidetoshi},
	title = {{Quantum annealing with antiferromagnetic fluctuations}},
	journal = {Phys. Rev. E},
	volume = {85},
	number = {5},
	pages = {051112},
	year = {2012},
	month = may,
	publisher = {American Physical Society},
	doi = {10.1103/PhysRevE.85.051112}
}

@article{serret2020,
  title = {Solving optimization problems with Rydberg analog quantum computers: Realistic requirements for quantum advantage using noisy simulation and classical benchmarks},
  author = {Serret, Michel Fabrice and Marchand, Bertrand and Ayral, Thomas},
  journal = {Phys. Rev. A},
  volume = {102},
  issue = {5},
  pages = {052617},
  numpages = {22},
  year = {2020},
  month = {Nov},
  publisher = {American Physical Society},
  doi = {10.1103/PhysRevA.102.052617},
  url = {https://link.aps.org/doi/10.1103/PhysRevA.102.052617}
}

@article{shin2014,
	author = {Shin, Seung Woo and Smith, Graeme and Smolin, John A. and Vazirani, Umesh},
	title = {{How "Quantum" is the D-Wave Machine?}},
	journal = {arXiv},
	year = {2014},
	month = jan,
	eprint = {1401.7087},
	doi = {10.48550/arXiv.1401.7087}
}

@article{scholl2021,
	author = {Scholl, Pascal and Schuler, Michael and Williams, Hannah J. and Eberharter, Alexander A. and Barredo, Daniel and Schymik, Kai-Niklas and Lienhard, Vincent and Henry, Louis-Paul and Lang, Thomas C. and Lahaye, Thierry and L{\ifmmode\ddot{a}\else\"{a}\fi}uchli, Andreas M. and Browaeys, Antoine},
	title = {{Quantum simulation of 2D antiferromagnets with hundreds of Rydberg atoms}},
	journal = {Nature},
	volume = {595},
	pages = {233--238},
	year = {2021},
	month = jul,
	issn = {1476-4687},
	publisher = {Nature Publishing Group},
	doi = {10.1038/s41586-021-03585-1}
}

@misc{scotti2024,
      title={A clustering aggregation algorithm on neutral-atoms and annealing quantum processors}, 
      author={Riccardo Scotti and Gabriella Bettonte and Antonio Costantini and Sara Marzella and Daniele Ottaviani and Stefano Lodi},
      year={2024},
      eprint={2412.07558},
      archivePrefix={arXiv},
      primaryClass={quant-ph},
      url={https://arxiv.org/abs/2412.07558}, 
}

@article{smith-miles2025,
doi = {10.1088/2058-9565/add61d},
url = {https://doi.org/10.1088/2058-9565/add61d},
year = {2025},
month = {may},
publisher = {IOP Publishing},
volume = {10},
number = {3},
pages = {033001},
author = {Smith-Miles, Kate A and Hoos, Holger H and Wang, Hao and Bäck, Thomas and Osborne, Tobias J},
title = {The travelling salesperson problem and the challenges of near-term quantum advantage},
journal = {Quantum Science and Technology}
}

@article{sprague2024,
	title = {Variational {Monte} {Carlo} with large patched transformers},
	volume = {7},
	issn = {2399-3650},
	url = {https://doi.org/10.1038/s42005-024-01584-y},
	doi = {10.1038/s42005-024-01584-y},
	number = {1},
	journal = {Communications Physics},
	author = {Sprague, Kyle and Czischek, Stefanie},
	month = {mar},
	year = {2024},
	pages = {90}
}

@article{srdinsek_2025,
  title = {Hybrid between biologically and quantum-inspired many-body states},
  author = {Srdin\ifmmode \check{s}\else \v{s}\fi{}ek, Miha and Waintal, Xavier},
  journal = {Phys. Rev. B},
  volume = {113},
  issue = {12},
  pages = {125107},
  numpages = {16},
  year = {2026},
  month = {Mar},
  publisher = {American Physical Society},
  doi = {10.1103/b2kz-15c5},
  url = {https://link.aps.org/doi/10.1103/b2kz-15c5}
}

@article{Stollenwerk2021,
  title={Quantum annealing applied to deconflicting optimal trajectories for air traffic management},
  author={Stollenwerk, Tobias and others},
  journal={IEEE Transactions on Intelligent Transportation Systems},
  volume={23},
  pages={16750--16763},
  year={2021},
  url={https://ieeexplore.ieee.org/stamp/stamp.jsp?arnumber=8643733}
}

@article{tarabunga2024,
	author = {Tarabunga, Poetri Sonya and Castelnovo, Claudio},
	title = {{Magic in generalized Rokhsar-Kivelson wavefunctions}},
	journal = {Quantum},
	volume = {8},
	pages = {1347},
	year = {2024},
	month = may,
	publisher = {Verein zur F{\ifmmode\ddot{o}\else\"{o}\fi}rderung des Open Access Publizierens in den Quantenwissenschaften},
	eprint = {2311.08463v2},
	doi = {10.22331/q-2024-05-14-1347}
}

@article{vandam2001,
	author = {van Dam, Wim and Mosca, Michele and Vazirani, Umesh},
	title = {{How Powerful is Adiabatic Quantum Computation?}},
	journal = {arXiv},
	year = {2002},
	month = jun,
	eprint = {quant-ph/0206003},
	doi = {10.1109/SFCS.2001.959902}
}

@article{Venturelli2022,
	author = {Venturelli, Davide and Kondratyev, Alexei},
	title = {{Reverse quantum annealing approach to portfolio optimization problems}},
	journal = {Quantum Mach. Intell.},
	volume = {1},
	number = {1},
	pages = {17--30},
	year = {2019},
	month = may,
	issn = {2524-4914},
	publisher = {Springer International Publishing},
	doi = {10.1007/s42484-019-00001-w}
}

@article{verstraete2006,
	author = {Verstraete, F. and Wolf, M. M. and Perez-Garcia, D. and Cirac, J. I.},
	title = {{Criticality, the Area Law, and the Computational Power of Projected Entangled Pair States}},
	journal = {Phys. Rev. Lett.},
	volume = {96},
	number = {22},
	pages = {220601},
	year = {2006},
	month = jun,
	publisher = {American Physical Society},
	doi = {10.1103/PhysRevLett.96.220601}
}

@article{vert2021,
	author = {Vert, Daniel and Sirdey, Renaud and Louise, St{\ifmmode\acute{e}\else\'{e}\fi}phane},
	title = {{Benchmarking Quantum Annealing Against {\textquotedblleft}Hard{\textquotedblright} Instances of the Bipartite Matching Problem}},
	journal = {SN Comput. Sci.},
	volume = {2},
	number = {2},
	pages = {106},
	year = {2021},
	month = apr,
	issn = {2661-8907},
	publisher = {Springer Singapore},
	doi = {10.1007/s42979-021-00483-1}
}

@article{vcerny1985,
	author = {{\ifmmode\check{C}\else\v{C}\fi}ern{\ifmmode\acute{y}\else\'{y}\fi}, V.},
	title = {{Thermodynamical approach to the traveling salesman problem: An efficient simulation algorithm}},
	journal = {J. Optim. Theory Appl.},
	volume = {45},
	number = {1},
	pages = {41--51},
	year = {1985},
	month = jan,
	issn = {1573-2878},
	publisher = {Kluwer Academic Publishers-Plenum Publishers},
	doi = {10.1007/BF00940812}
}

@article{villain1980,
	author = {Villain, J. and Bidaux, R. and Carton, J.-P. and Conte, R.},
	title = {{Order as an effect of disorder}},
	journal = {J. Phys.},
	volume = {41},
	number = {11},
	pages = {1263--1272},
	year = {1980},
	month = nov,
	issn = {0302-0738},
	publisher = {Soci{\ifmmode\acute{e}\else\'{e}\fi}t{\ifmmode\acute{e}\else\'{e}\fi} Fran{\ifmmode\mbox{\c{c}}\else\c{c}\fi}aise de Physique},
	doi = {10.1051/jphys:0198000410110126300}
}

@article{white1992,
  title = {Density matrix formulation for quantum renormalization groups},
  author = {White, Steven R.},
  journal = {Phys. Rev. Lett.},
  volume = {69},
  issue = {19},
  pages = {2863--2866},
  numpages = {0},
  year = {1992},
  month = {Nov},
  publisher = {American Physical Society},
  doi = {10.1103/PhysRevLett.69.2863},
  url = {https://link.aps.org/doi/10.1103/PhysRevLett.69.2863}
}

@article{wolf2008,
  title = {Area Laws in Quantum Systems: Mutual Information and Correlations},
  author = {Wolf, Michael M. and Verstraete, Frank and Hastings, Matthew B. and Cirac, J. Ignacio},
  journal = {Phys. Rev. Lett.},
  volume = {100},
  issue = {7},
  pages = {070502},
  numpages = {4},
  year = {2008},
  month = {Feb},
  publisher = {American Physical Society},
  doi = {10.1103/PhysRevLett.100.070502},
  url = {https://link.aps.org/doi/10.1103/PhysRevLett.100.070502}
}

@article{wu2024,
	author = {Wu, Dian and Rossi, Riccardo and Vicentini, Filippo and Astrakhantsev, Nikita and Becca, Federico and Cao, Xiaodong and Carrasquilla, Juan and Ferrari, Francesco and Georges, Antoine and Hibat-Allah, Mohamed and Imada, Masatoshi and L{\ifmmode\ddot{a}\else\"{a}\fi}uchli, Andreas M. and Mazzola, Guglielmo and Mezzacapo, Antonio and Millis, Andrew and Moreno, Javier Robledo and Neupert, Titus and Nomura, Yusuke and Nys, Jannes and Parcollet, Olivier and Pohle, Rico and Romero, Imelda and Schmid, Michael and Silvester, J. Maxwell and Sorella, Sandro and Tocchio, Luca F. and Wang, Lei and White, Steven R. and Wietek, Alexander and Yang, Qi and Yang, Yiqi and Zhang, Shiwei and Carleo, Giuseppe},
	title = {{Variational benchmarks for quantum many-body problems}},
	journal = {Science},
	volume = {386},
	number = {6719},
	pages = {296--301},
	year = {2024},
	month = oct,
	issn = {0036-8075},
	publisher = {American Association for the Advancement of Science},
	doi = {10.1126/science.adg9774}
}

@article{yarkoni2022,
doi = {10.1088/1361-6633/ac8c54},
url = {https://doi.org/10.1088/1361-6633/ac8c54},
year = {2022},
month = {sep},
publisher = {IOP Publishing},
volume = {85},
number = {10},
pages = {104001},
author = {Yarkoni, Sheir and Raponi, Elena and Bäck, Thomas and Schmitt, Sebastian},
title = {Quantum annealing for industry applications: introduction and review},
journal = {Reports on Progress in Physics}
}

@article{young2010,
  title = {First-Order Phase Transition in the Quantum Adiabatic Algorithm},
  author = {Young, A. P. and Knysh, S. and Smelyanskiy, V. N.},
  journal = {Phys. Rev. Lett.},
  volume = {104},
  issue = {2},
  pages = {020502},
  numpages = {4},
  year = {2010},
  month = {Jan},
  publisher = {American Physical Society},
  doi = {10.1103/PhysRevLett.104.020502},
  url = {https://link.aps.org/doi/10.1103/PhysRevLett.104.020502}
}

@article{zener1932,
	author = {Zener, Clarence},
	title = {{Non-adiabatic crossing of energy levels}},
	journal = {Proc. R. Soc. London A},
	volume = {137},
	number = {833},
	pages = {696--702},
	year = {1932},
	month = sep,
	issn = {0950-1207},
	publisher = {The Royal Society},
	doi = {10.1098/rspa.1932.0165}
}
